\documentclass[aps,prc,twocolumn,showpacs,floatfix,preprintnumbers,%
superscriptaddress,amsmath,amssymb,longbibliography]{revtex4-1}

\usepackage{tabularx}
\usepackage{mathtools}
\usepackage{pifont}
\usepackage[colorlinks,citecolor=blue]{hyperref}
\usepackage[colorinlistoftodos]{todonotes}
\usepackage[normalem]{ulem}
\usepackage{color}
\usepackage[utf8]{inputenc}
\usepackage{bm}
\usepackage{caption}
\usepackage{subcaption}

\makeatletter
\def\@bibdataout@aps{%
\immediate\write\@bibdataout{%
@CONTROL{%
apsrev41Control%
\longbibliography@sw{%
    ,author="08",editor="1",pages="1",title="0",year="1"%
    }{%
    ,author="08",editor="1",pages="1",title="",year="1"%
    }%
  }%
}%
\if@filesw \immediate \write \@auxout {\string \citation {apsrev41Control}}\fi 
}
\makeatother

\graphicspath{{./}{figures/}}

\definecolor{pastelgray}{rgb}{0.81, 0.81, 0.77}
\definecolor{beaublue}{rgb}{0.9, 0.9, 0.93}

\begin{document}
\title{I-C-Q relations for rapidly rotating stable hybrid stars}

\author{Sujan Kumar Roy}
\affiliation{Physics Group, Variable Energy Cyclotron Centre, Kolkata 700064, India }
\affiliation{Homi Bhabha National Institute, Mumbai 400094, India }
\author{Gargi Chaudhuri}
\affiliation{Physics Group, Variable Energy Cyclotron Centre, Kolkata 700064, India }
\affiliation{Homi Bhabha National Institute, Mumbai 400094, India }
\date{\today}

\begin{abstract}
A number of hadronic equations of state for neutron stars have been investigated for the purpose of the present paper, considering the fact that at sufficiently high density, heavy baryons and quark phases may appear. The observational limits from NICER, GW170817, etc., are obeyed by our choice of equations of state. The universal relations are investigated for both slowly and rapidly rotating neutron stars with heavy baryons present inside the core. For slowly rotating stars, the universality of the I-Love-Q relations is verified, and the I-C-Q relations are inferred to be universal for rapidly rotating stars. Further, we extend the investigation to obtain the universal relations for compact stars containing the quark core, where the connected stable branch of such hybrid stars is considered. The parameters of the I-Love-Q and I-C-Q universal relations are obtained for slowly rotating and rapidly rotating hybrid stars, respectively. These relations would enable extracting information, within the context of general relativity, from astrophysical systems involving rapidly rotating neutron stars.
\end{abstract}
\maketitle

\section{Introduction}
\label{intro}
Neutron stars (NSs) are very compact objects and natural laboratories for studying the physics of matters at extreme densities. The equation of state (EoS) describes compact object matter under extreme conditions, where exotic phases of matter may also appear \cite{Lattimer_2012_DkuaX, Lattimer_2016_DVZWU, _zel_2016_MkmxI, Oertel_2017_NDBAA, Baym_2018_fVTXR, Tolos_2020_aiFBV, Sedrakian_2023_UJvrR}. EoS plays a significant role in NS study as the macroscopic properties of the compact objects, e.g., mass, radius, etc., are sensitive to the nature of the EoS \cite{Lattimer_2001_WTceP, Ji_2019_aOrbX}. Therefore, measuring these quantities provides valuable insights into the extreme conditions within neutron stars and the nature of dense matter physics.

X-ray pulse profiling of NICER has added a new dimension to the pulsar observations, which includes the mass, radius, and surface composition of NSs. Although the physics of matter at supranuclear density is poorly understood, these measurements help gain further insight into the nature of the EoS \cite{Xie_2020_QhQne}. For instance, Fonseca et al. \cite{Fonseca_2021_SOyMc} reported the mass of PSR J0740+6620 to be 2.08$^{+0.07} _{-0.07} M_\odot$ within 68.3\% credibility interval. The mass determination of the massive pulsar provides important constraints on the EoS. However, the inconclusive radius of the mentioned pulsar restricts our inferences. Recent NICER observations \cite{Miller_2019_dFNjM, Riley_2019_oQUYh, Miller_2021_pAcGM, Riley_2021_kbhlr, Salmi_2022_wcauZ} determine the radius of the NS with acceptable accuracy, which adds further to the EoS study.

In modelling NSs, EoS plays a crucial role as the properties of the compact objects are determined by the EoS. The parameters like mass, radius and higher multipole moments are responsible for determining the structure of the space-time surrounding the compact objects. Therefore, it is further possible to put constraints on the EoS due to inferences of mass and tidal deformability drawn from events like GW170817 \cite{Bauswein_2017_JoXvH, De_2018_LcNIX, Malik_2018_qAfAT, Most_2018_JHlGX, Tews_2018_VRpfs}. From particle physics point of view, several EoSs, which reproduce the bulk properties of finite nuclei, are consistent with the astrophysical constraints as well \cite{Hebeler_2010_qcfiw, Oertel_2017_NDBAA, Dutra_2014_LtrKV, Tews_2018_cbmyx, Malik_2018_qAfAT}. Interestingly, it turns out that for these EoSs, the relations between various NS parameters—such as the moment of inertia, quadrupole moment, and tidal deformability—are found to be quite insensitive to microscopic variations in the pressure-energy density relation \cite{Yagi_2013_cPzvP, Maselli_2013_KEJBr}. 

The relations between higher multipole moments, known as the universal relations, are almost EoS independent, which makes neutron stars promising candidates alongside black holes (BHs) for testing theories of gravity. Therefore, the universality relations between various NS observables hold particular importance in the literature \cite{Yagi_2013_cPzvP, Maselli_2013_KEJBr, Yagi_2013_yxUgU, Urbanec_2013_wNmiD, Baub_ck_2013_gJPHq, Doneva_2013_8E7jv, Cipolletta_2015_iJrxY}. For instance, the relation between the moment of inertia $I$, the Love number $\overline{\Lambda}$ and the quadrupole moment $Q$, i.e., I-Love-Q relation is approximately universal for non-rotating, slowly rotating non-magnetized stars within small tidal deformation approximation \cite{Yagi_2013_yxUgU, Urbanec_2013_wNmiD}. NSs, however, exhibit magnetic fields ranging from 10$^8$ to 10$^{15}$ gauss, making them some of the most magnetized objects in the universe \cite{Phinney_1994_TmSXL, Manchester_2004_tPdWU, Thompson_1995_tjhSK, Kaspi_2004_nTcSP, Mereghetti_2008_NCHAj, Makishima_1999_DRwlJ}. Therefore, efforts have been made to explore universal relations for moderately high magnetic fields in NSs \cite{Haskell_2013_cKUtF}. The relevance of tidal effects in determining the inspiral gravitational wave (GW) signal for a coalescing binary NS is discussed in \cite{Dudi_2018_jaQnY}. This relevance invokes the relaxation of the adiabatic approximation \cite{Maselli_2012_WrTFq, Gagnon_Bischoff_2018_FCGMY, Castro_2021_xmbve}, and extends the exploration of universal relations, such as the relations between $I$, compactness $C$, and $Q$, i.e., I-C-Q relation. Moreover, the observational evidence of spin variations in pulsars, the spin-period ranging from ms to few seconds \cite{Manchester_2005_rjESW, Liu_2022_BdFHo}, justifies the extension of universal relations to rapidly rotating stars \cite{Doneva_2013_8E7jv, Pappas_2014_TIiDy,  Chakrabarti_2014_kBsdN, Sun_2020_QTRnD, Khosravi_Largani_2022_cvIJr}.

The heavy baryons, including hyperons and $\Delta$-admixed matter, give rise to a softer EoS, affecting fundamental NS properties such as maximum masses of static NSs, their radii, and tidal deformabilities \cite{Weber_2005_SYDRw, Tolos_2016_kVYCR}. Recent studies have quantitatively demonstrated that the presence of hyperons can lower the maximum NS mass while also affecting the stiffness of the EoS, resulting in higher tidal deformability \cite{Bednarek_2012_dOXAn, Logoteta_2021_MkalM}. On the other hand, the hadron-quark phase transition is a key factor in determining the bulk properties of NSs. This phase transition significantly affects the EoS at high densities, impacting the maximum mass of static NSs, their radii, tidal deformabilities, core-collapse and GWs from NS merger events \cite{Burgio_2002_gVOwJ, Hanauske_2019_xvUXm, Jakobus_2022_rTUic}.

The present work aims to assess various universal relations applicable to static and rotating NSs, employing EoSs incorporating heavy baryonic degrees of freedom at supranuclear densities at zero temperature. The inclusion of heavy baryonic degrees of freedom in the NS model allows us to explore a range within the pressure-energy density plane. Several studies are already available in the literature which explore in detail the almost EoS-independent relations between different integral quantities of compact stars \cite{Yagi_2013_cPzvP, Yagi_2013_yxUgU, Maselli_2013_KEJBr, Baub_ck_2013_gJPHq, Pappas_2014_TIiDy, Chakrabarti_2014_kBsdN, Chatterjee_2016_ijIxE, Marques_2017_dQf8a, Li_2018_yrEXW, Lenka_2019_VAoni, Raduta_2020_YKkHx, Li_2023_wnUwG}. Several studies have focused on deriving various quasi-universal relations by utilizing a variety of EoS models \cite{Urbanec_2013_wNmiD, Marques_2017_dQf8a, Raduta_2020_YKkHx, Suleiman_2021_lhDZ3, Khosravi_Largani_2022_cvIJr, Yang_2022_RbVSz, Saes_2022_KdfpG, Carlson_2023_0TLyw, Suleiman_2024_7cKmd}. The EoS insensitive relations are also useful for testing theories of gravity \cite{Pappas_2019_zqY5v, Sham_2014_EC3zI, Doneva_2016_QRTMQ}. Some studies have attempted to derive universal relations using more generalized parametric and phenomenological EoSs within the context of general relativity \cite{Godzieba_2021_UBDLe, Saes_2022_KdfpG, Legred_2024_WRfSP, Suleiman_2024_7cKmd}. However, most of these studies are limited to the slow-rotation approximation. Attempts have been made to obtain quasi-universal relations for rapidly rotating stars as well \cite{Doneva_2013_8E7jv, Pappas_2014_TIiDy, Chakrabarti_2014_kBsdN, Breu_2016_WqIRD, Khosravi_Largani_2022_cvIJr}. A few other studies are dedicated to exploring the role of heavy baryons in determining the properties of NSs \cite{Bednarek_2012_dOXAn, Chatterjee_2016_ijIxE, Li_2018_yrEXW, Lenka_2019_VAoni, Li_2020_il2wz, Dexheimer_2021_Nc522, Logoteta_2021_MkalM, Sedrakian_2022_tiZFB} and establishing universal relations in their presence, such as the I-Love-Q relation for slowly rotating NSs and I-C-Q relation for rapidly rotating NSs \cite{Marques_2017_dQf8a, Lenka_2019_VAoni, Raduta_2020_YKkHx, Li_2023_wnUwG}. First, we obtain several quasi-universal relations for both slowly and rapidly rotating NSs, considering heavy baryon degrees of freedom inside the core. Comparison with earlier results available in the literature suggests that our results fall within the acceptable range.

Within the context of current astrophysical observations, several papers discuss the hadron-quark phase transition in the cores of compact stars (e.g., refs. \cite{Bonanno_2012_Ca0lZ, Ayvazyan_2013_NZ1QJ, Pereira_2018_LsPLH, Nandi_2018_YZU8D, Rather_2021_v7Q9O, Li_2021_e1GSh}). Further, EoS insensitive relations for quark stars and hybrid stars are discussed by several authors \cite{Urbanec_2013_wNmiD, Yagi_2013_cPzvP, Yagi_2014_lZhJS, Yagi_2017_uWwLw, Bozzola_2019_861n9, Yeung_2021_dc0JE, Khosravi_Largani_2022_cvIJr}. For instance, within the framework of slow-rotation approximation, the relations between the moment of inertia, angular momentum parameter and quadrupole moments obtained for strange stars differ from those of NSs \cite{Urbanec_2013_wNmiD}. Rapidly rotating quark stars and their quasi-universal relations are considered for investigation in refs \cite{Yagi_2013_cPzvP, Yagi_2014_lZhJS, Yagi_2017_uWwLw}. Ref \cite{Yagi_2014_lZhJS}, described the quark matter within the context of refs \cite{Lattimer_2001_WTceP, Farhi_1984_dqFvA}, shows the variability of the relations for quark stars to be $\sim$ 10\%. Bozzola et al. (2019) extended the study for hybrid stars to investigate the relations between angular momentum, rest mass and gravitational mass for turning-point models \cite{Bozzola_2019_861n9}. Authors in refs \cite{Yeung_2021_dc0JE, Burikham_2022_Kq9sd} discuss the validity of I-Love-Q relations for HSs within slow-rotation approximation. Ref \cite{Khosravi_Largani_2022_cvIJr} considered hadrons within mean-field models and quarks within MIT Bag model \cite{Fischer_2011_DbzLV}, vector-interaction-enhanced bag model \cite{Kl_hn_2015_lo3ys}, String-flip model \cite{Bastian_2018_K1F9o} and NJL model \cite{Klevansky_1992_2oEgq, Beni__2015_PdKX3, Baym_2018_fVTXR}) to investigate the universality between the maximum mass and the radius of non-rotating and maximally rapidly rotating configurations, as well as the I-C relation. A similar kind of study has recently been performed with a machine-learning technique to obtain relations for rapidly rotating HSs \cite{Papigkiotis_2023_a8kst}. Thus, a systematic study of the universal relations for the most general case of core composition---dense matter with a full baryon octet and hadron-quark phase transition---considering arbitrary rotation has yet to be added to the literature. To address this, we construct a set of stable hybrid stars \cite{Alford_2013_7wL5W} from our chosen set of EoSs and obtain the quasi-universal relations between the moment of inertia, compactness, quadrupole moments, etc., for rapidly rotating hybrid stars. These quasi-universal relations for the rapidly rotating sequences of hybrid stars are presented to add to the set of other neutron star universal relations in the literature. Finally, we compare the relations we obtained with others in the literature to address the role of core composition variability in determining quasi-universal relations.

We introduce our chosen set of EoS models in section \ref{eos_model} \cite{obspmHome, doiCompOSEReference, Typel_2015_gH0QU}. The description of equilibrium configurations for slowly rotating to rapidly rotating stars is given in section \ref{equilibrium_model}. The obtained quasi-universal relations, in conjunction with a comparison to earlier results, are described in section \ref{result_discussion}. Finally, we summarize our work and present concluding remarks in the last section \ref{conclusion}.

\twocolumngrid
\begin{figure}
  \centering
  \captionsetup{justification=raggedright}
  \subfloat[Pressure vs energy density]{\includegraphics[width=1.0\linewidth]{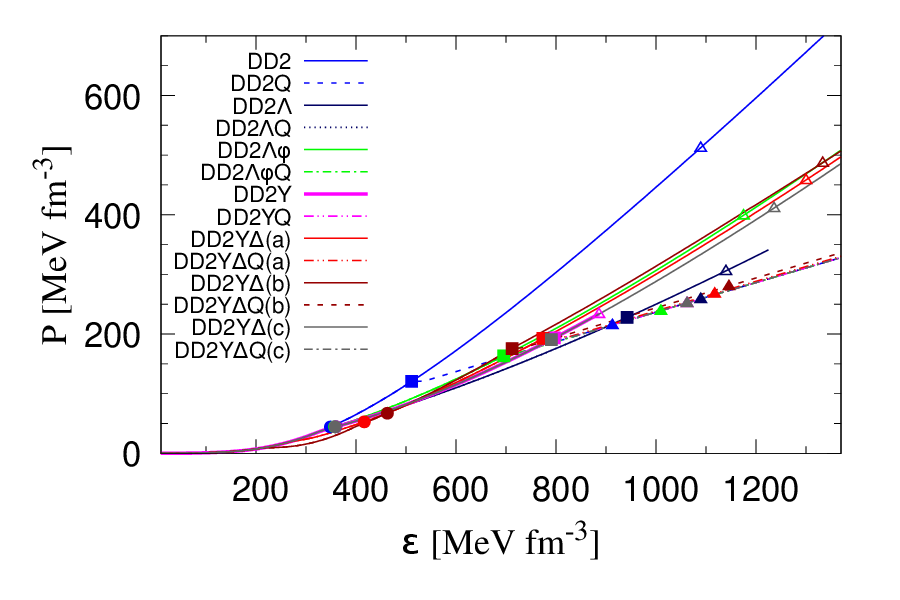}}\hfill
  \subfloat[$c_s^2$ vs energy density]{\includegraphics[width=1.0\linewidth]{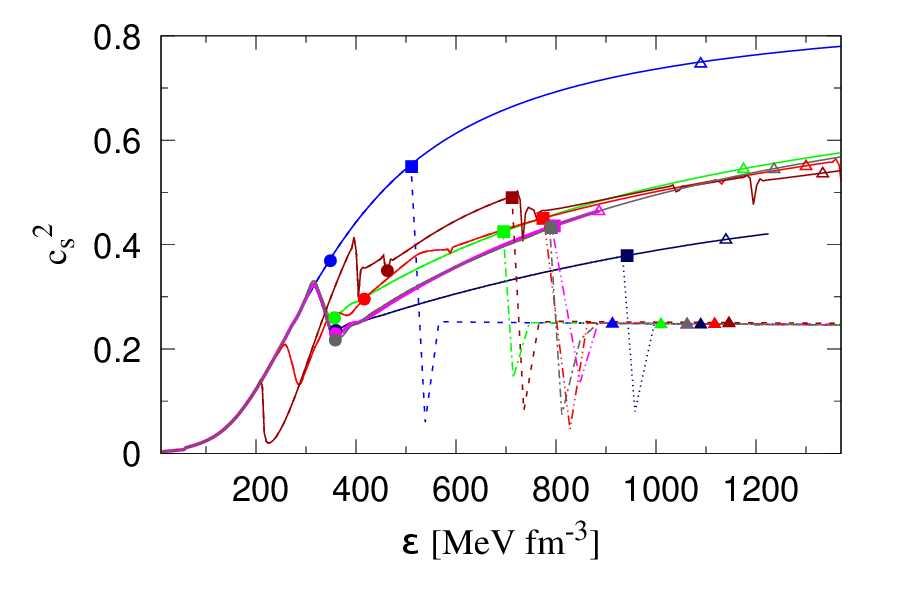}}
\caption{NS EoSs considering different compositions are presented in (a). The solid and dashed curves represent hadronic EoSs and quark admixed hadronic EoSs, respectively. Associated speed of sound squared, $c_s^2$ for the considered EoSs are also presented in (b). The solid circles are the EoS points which lead to 1.4M${_\odot}$ configurations; solid squares represent the point of phase transition, the triangular points lead to maximum NS mass while considering only hadronic core, whereas solid triangular points represent the maximum mass of static configurations with quark core.}
\label{fig1_eos} 
\end{figure}

\begin{figure}
  \centering
  \captionsetup{justification=raggedright}
  \subfloat[Static limit of M-R]{\includegraphics[width=1.0\linewidth]{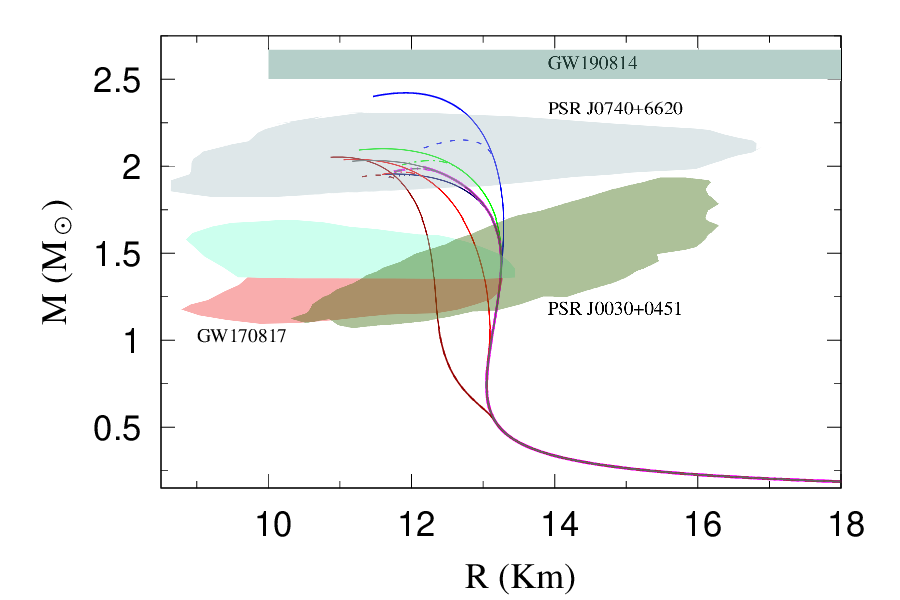}}\hfill
  \subfloat[Keplerian limit of M-R]{\includegraphics[width=1.0\linewidth]{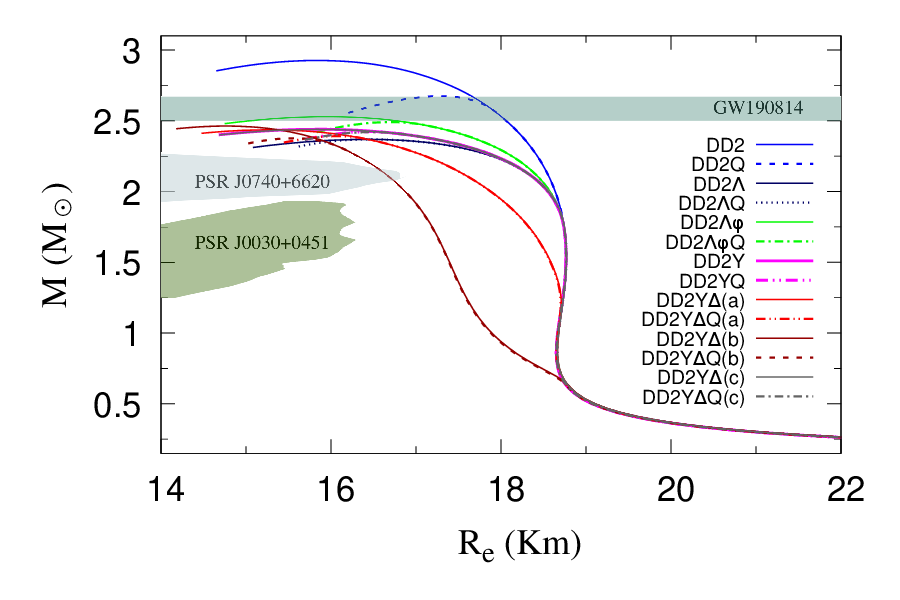}}
  \caption{ Mass-radius relations of NSs in the static and Keplerian limits for various EoSs are presented with NICER measurements of PSR J0030+0451 \cite{Miller_2019_dFNjM, Riley_2019_oQUYh}, and PSR J0740+6620 \cite{Miller_2021_pAcGM, Riley_2021_kbhlr} (95\% credible interval). The NS radii constraints due to GW170817 \cite{Abbott_2018_EX9Nk} (in subfigure (a)) and the extracted mass range of GW190814's secondary \cite{Abbott_2020_jIBsD} are also shown. Color code is the same for both (a) and (b).}
  \label{fig2_M-R}
\end{figure}

\begin{figure}
  \centering
  \captionsetup{justification=raggedright}
  {\includegraphics[width=1.0\linewidth]{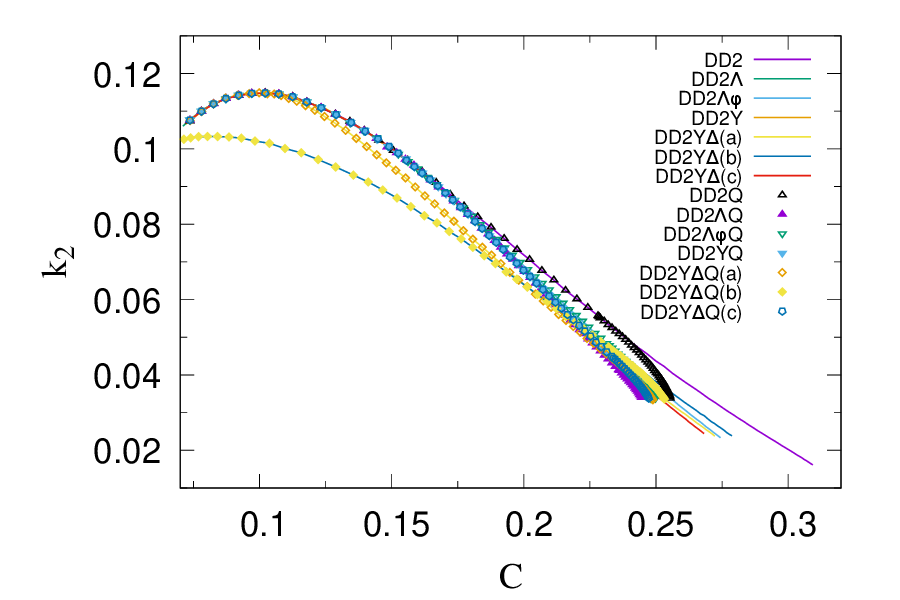}}
  \caption{The dimensionless tidal Love number $k_2$ as a function of compactness C for both hadronic and hadron-quark hybrid EoSs.}
  \label{fig3_k2-C}
\end{figure}

\begin{figure}
  \centering
  \captionsetup{justification=raggedright}
  \subfloat{\includegraphics[width=1.0\linewidth]{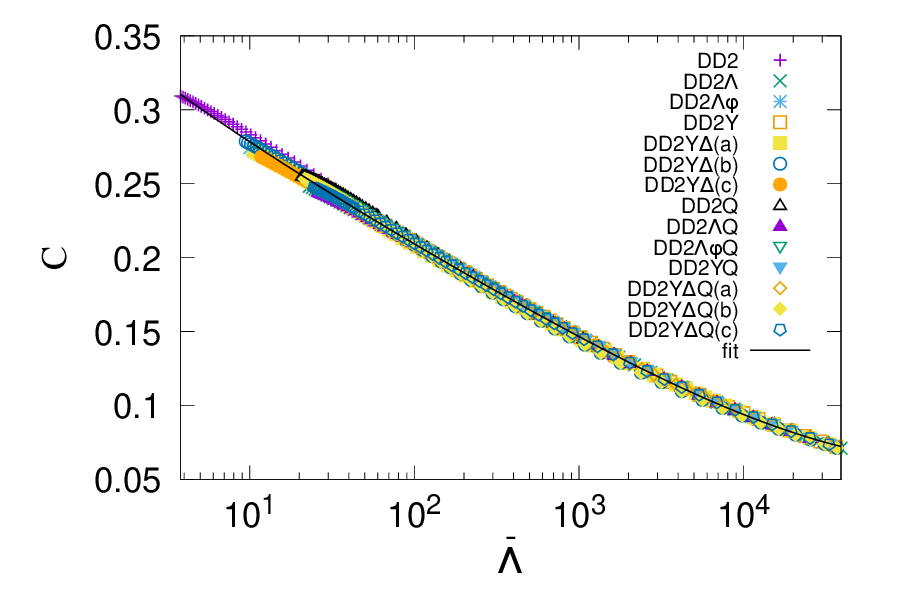}}\hfill
  \vspace{-0.25cm}
  \subfloat{\includegraphics[width=1.0\linewidth]{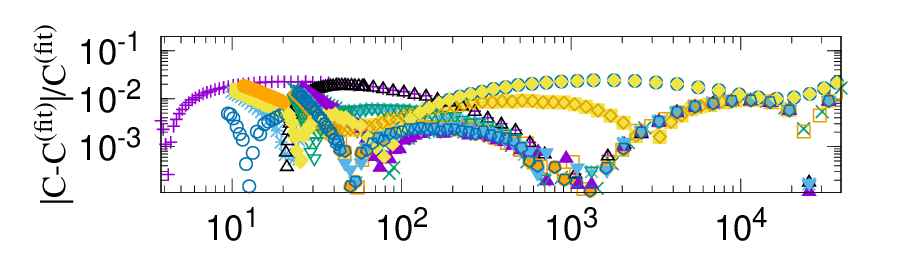}}
  \caption{ The C-$\overline{\Lambda}$ relation using hadronic and hybrid EoSs. In the upper panel, calculated C-$\overline{\Lambda}$ points are presented along with the fitting curve. Fractional errors between the calculation and fitting curve are in the bottom panel.}
  \label{fig4_C-L}
\end{figure}

\begin{figure}
  \centering
  \captionsetup{justification=raggedright}
  \subfloat{\includegraphics[width=1.0\linewidth]{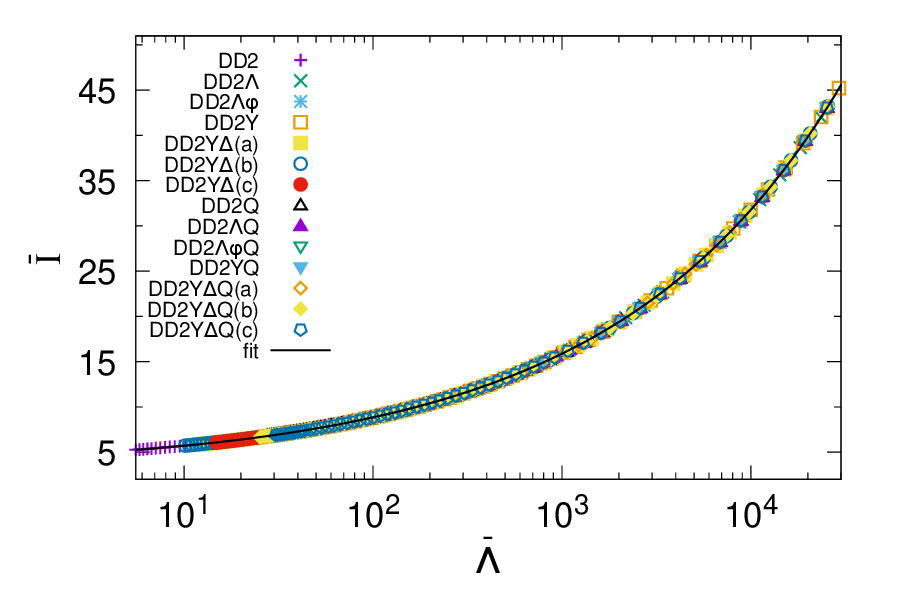}}\hfill
  \vspace{-0.25cm}
  \subfloat{\includegraphics[width=1.0\linewidth]{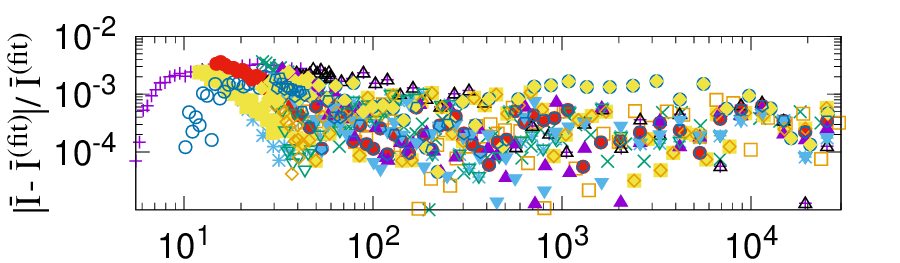}}
  \caption{$\overline{I}$-$\overline{\Lambda}$ relation is presented for slowly rotating NSs and HSs. The calculated $\overline{I}$-$\overline{\Lambda}$ points and the fitting curve for all the EoSs are presented in the upper panel. The lower panel presents the fractional errors between fitting and numerical calculations.}
  \label{fig5_I-L_sr}
\end{figure}

\begin{figure}
  \centering
  \captionsetup{justification=raggedright}
  \subfloat{\includegraphics[width=1.0\linewidth]{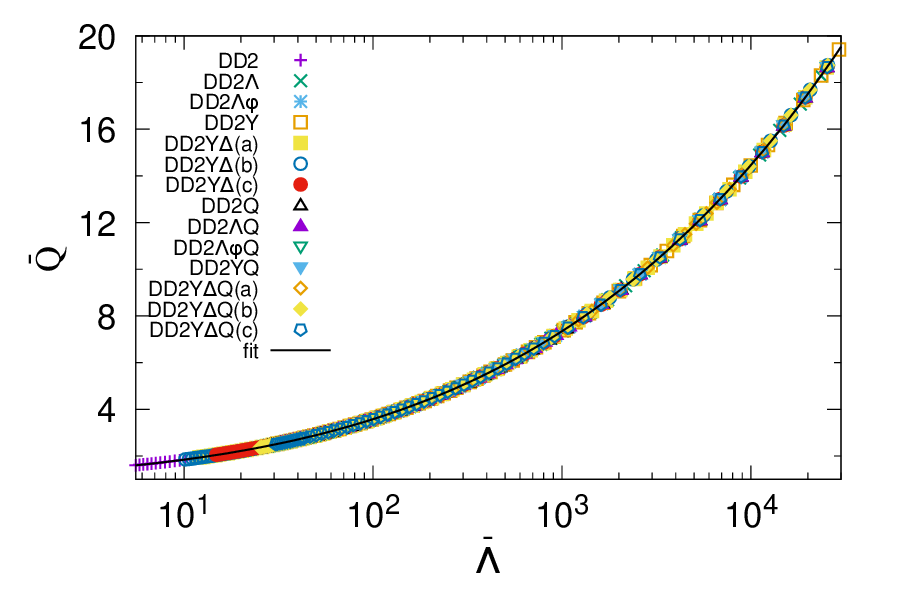}}\hfill
  \vspace{-0.25cm}
  \subfloat{\includegraphics[width=1.0\linewidth]{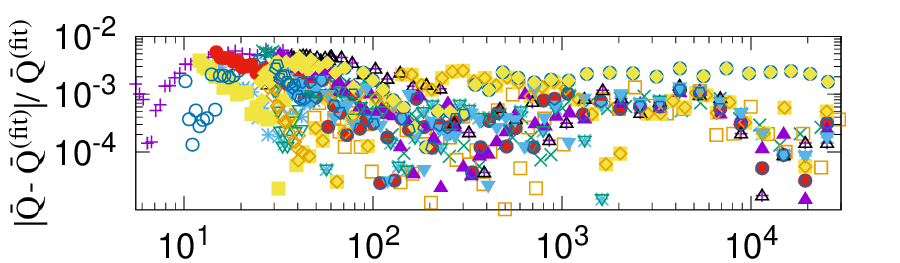}}
  \caption{$\overline{Q}$-$\overline{\Lambda}$ relation (top) along with the deviation (bottom) is presented for slowly rotating NSs and HSs.}
  \label{fig7_Q-L_sr}
\end{figure}

\begin{figure}
  \centering
  \captionsetup{justification=raggedright}
  \subfloat{\includegraphics[width=1.0\linewidth]{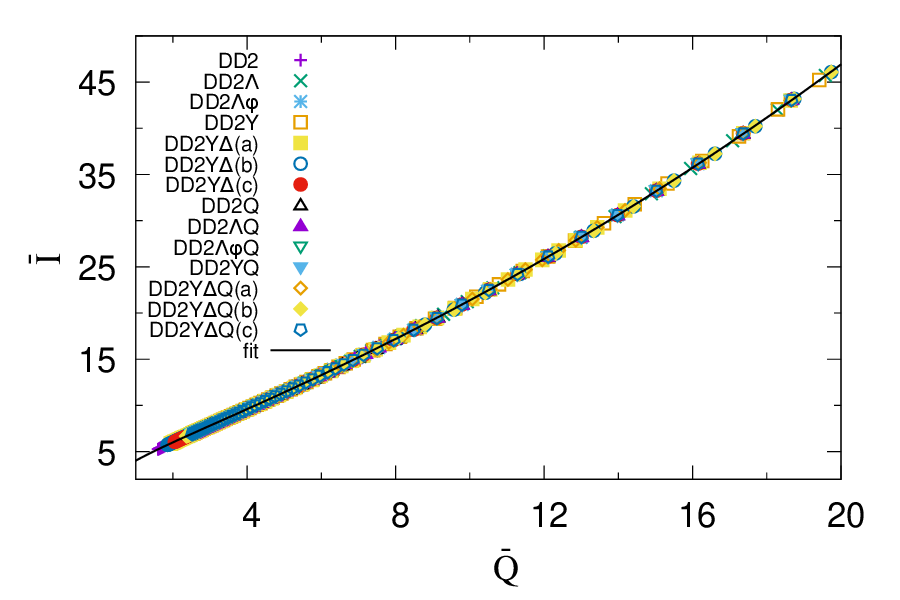}}\hfill
  \vspace{-0.25cm}
  \subfloat{\includegraphics[width=1.0\linewidth]{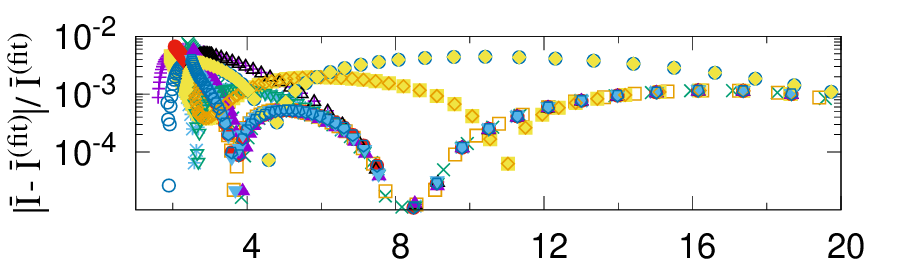}}
  \caption{$\overline{I}$-$\overline{Q}$ relation (top) along with the deviation (bottom) is presented for slowly rotating NSs and HSs.}
  \label{fig6_I-Q_sr}
\end{figure}

\begin{figure}
  \centering
  \captionsetup{justification=raggedright}
  {\includegraphics[width=1.0\linewidth]{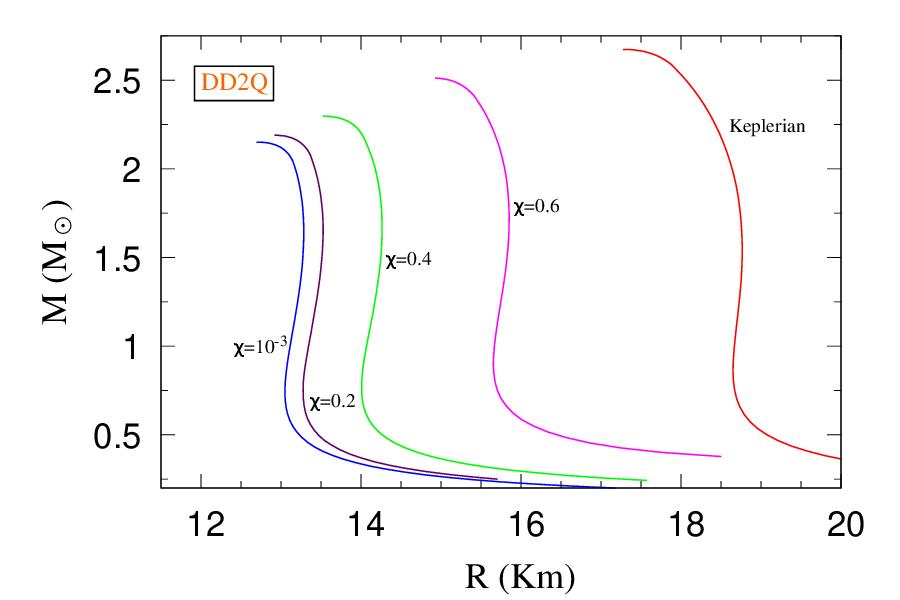}}
  \caption{Mass-radius relation of HSs for spin parameter $\chi =$ 10$^{-3}$, 0.2, 0.4, and 0.6 along with Keplerian sequence using DD2Q EoS.}
  \label{fig2a_M-R}
\end{figure}

\begin{figure}
  \centering
  \captionsetup{justification=raggedright}
  \subfloat{\includegraphics[width=1.0\linewidth]{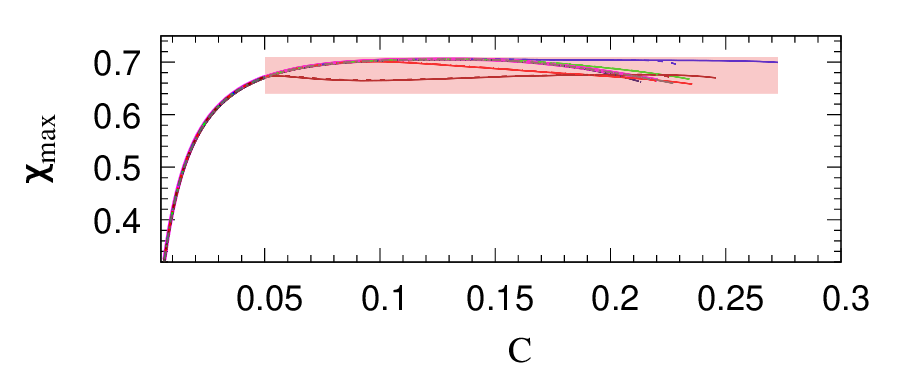}}\hfill
  \vspace{-0.2cm}
  \subfloat{\includegraphics[width=1.0\linewidth]{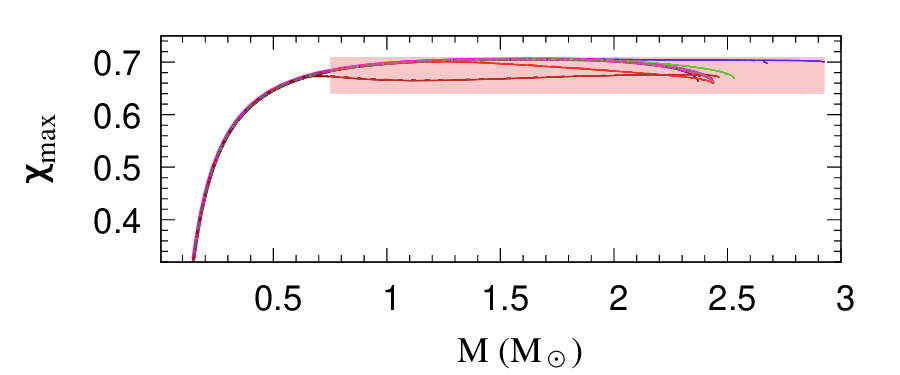}}
  \caption{ The maximum spin parameter $\chi_{\text{max}}$ corresponds to the Keplerian sequence as a function of compactness and gravitational mass of the star, respectively.}
  \label{fig2b_chi}
\end{figure}

\begin{figure}
  \centering
  \captionsetup{justification=raggedright}
  \subfloat{\includegraphics[width=1.0\linewidth]{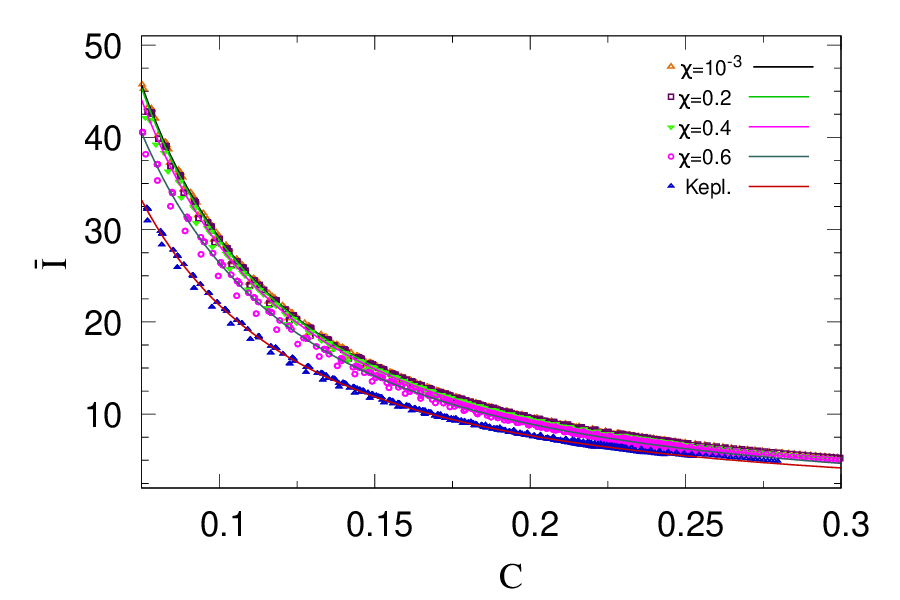}}\hfill
  \vspace{-0.25cm}
  \subfloat{\includegraphics[width=1.0\linewidth]{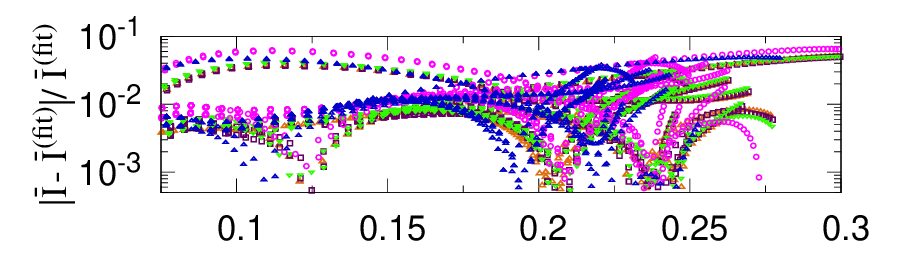}}
  \caption{$\overline{I}$-C relation (top) along with the deviation (bottom) is presented for rapidly rotating NSs and HSs for various spin sequences $\chi =$ 10$^{-3}$, 0.2, 0.4, 0.6, and the Keplerian sequence.}
  \label{fig8_I-C}
\end{figure}

\begin{figure}
  \centering
  \captionsetup{justification=raggedright}
  \subfloat{\includegraphics[width=1.0\linewidth]{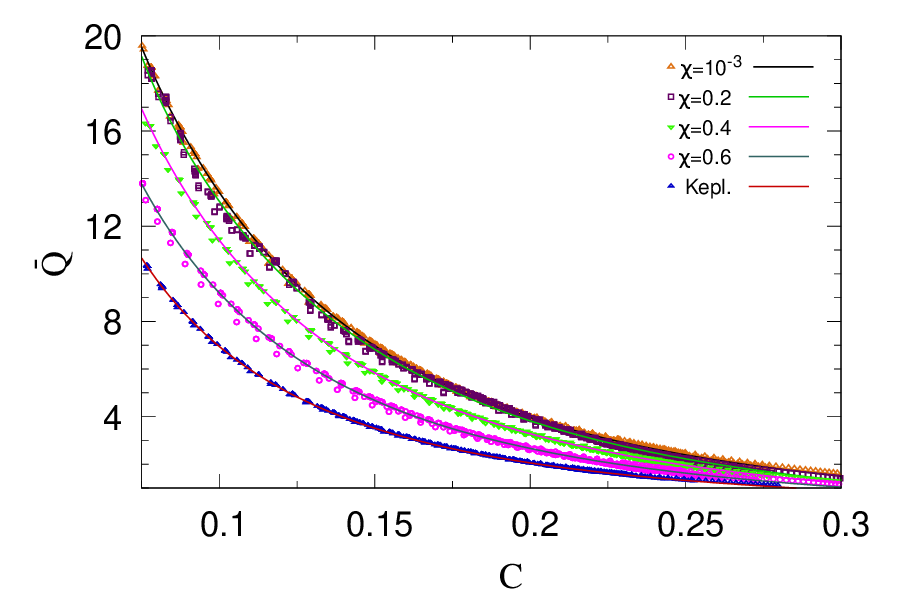}}\hfill
  \vspace{-0.25cm}
  \subfloat{\includegraphics[width=1.0\linewidth]{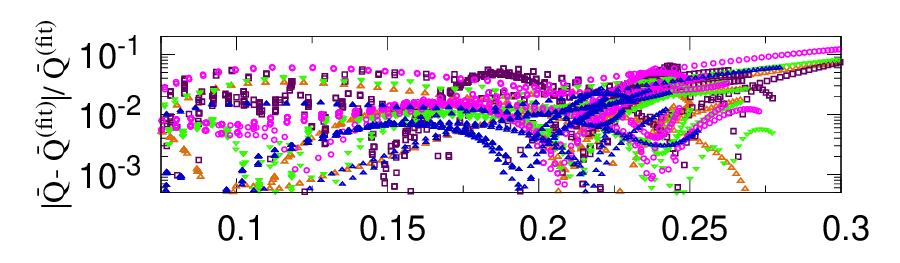}}
  \caption{$\overline{Q}$-C relation (top) along with the deviation (bottom) is presented for rapidly rotating NSs and HSs for various spin sequences ($\chi = $) 0.2, 0.4, 0.6, and the Keplerian sequence.}
  \label{fig9_Q-C}
\end{figure}

\begin{figure}
  \centering
  \captionsetup{justification=raggedright}
  \subfloat{\includegraphics[width=1.0\linewidth]{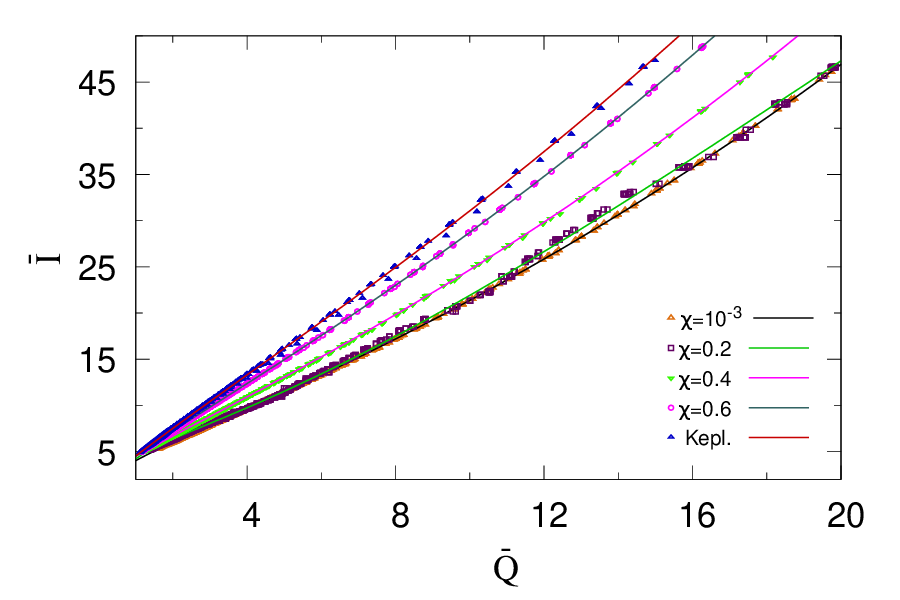}}\hfill
  \vspace{-0.25cm}
  \subfloat{\includegraphics[width=1.0\linewidth]{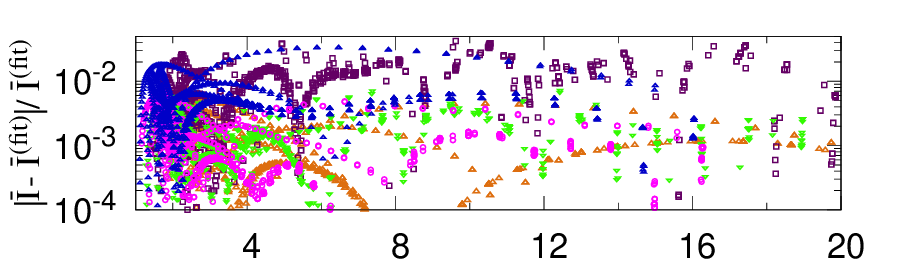}}
  \caption{$\overline{I}$-$\overline{Q}$ relation (top) along with the deviation (bottom) is presented for rapidly rotating NSs and HSs for various spin sequences ($\chi = $) 0.2, 0.4, 0.6, and the Keplerian sequence.}
  \label{fig10_I-Q}
\end{figure}

\section{Equation of State}
\label{eos_model}
In this study, we employ the nuclear statistical model to describe an ensemble of interacting nucleons at zero temperature. Additionally, we use the Relativistic Mean Field (RMF) model, which incorporates the exchange interactions of the isoscalar–scalar $\sigma$, the isoscalar–vector $\omega$, and the isovector–vector $\rho$ mesons, as detailed in Ref \cite{Typel_2010_CZbbi}. Then, within this formalism, we consider the appearance of hyperons, $\Delta$-admixed matter at sufficiently high densities. For the present calculation, we obtain the EoS tables from the publicly available CompOSE database \cite{obspmHome, doiCompOSEReference, Oertel_2017_NDBAA, Typel_2015_gH0QU} to describe nuclear matter and matter with heavy baryon degrees of freedom.

\subsection{Heavy Baryon Admixed Nuclear Matter}
\label{eos_nuclear}
We use the EoS developed in ref. \cite{Hempel_2010_mRz7r} for an ensemble of interacting nucleons for nucleonic degrees of freedom at zero temperature. The EoS model is referred to as 'DD2' in original works \cite{Hempel_2010_mRz7r, Typel_2010_CZbbi}, and here we adopt this designation. The model is further extended to account for the contribution of $\Lambda$ hyperons at high densities by ref. \cite{Banik_2014_QOKgS}. The EoS table, referred to as 'DD2$\Lambda$', considers attractive potential depth for $\Lambda$ hyperons. Therefore, the onset of $\Lambda$ hyperons softens the EoS, which results in a reduction in the maximum mass of the static NS sequence as compared to the 'DD2' model. The term maximum mass would refer, in general, to the static mass of the last stable configuration unless otherwise mentioned. On the other hand, the EoS table 'DD2$\Lambda \phi$' taken from the same work accounts for the repulsive interaction between $\Lambda$ hyperons mediated by the strange $\phi$ meson. This repulsive $\Lambda-\Lambda$ interaction results in a slight increase in the maximum mass. Different interactions' parameters and resulting NS sequences are outlined in the original works \cite{Hempel_2010_mRz7r, Typel_2010_CZbbi, Banik_2014_QOKgS}. It should be noted that refs \cite{Fortin_2017_jzm5q, Fortin_2020_GMbFF} developed EoSs for NSs, which also treat $\Sigma$, $\Lambda$ degrees of freedom including with $\phi$ mesons. However, ref \cite{Fortin_2020_GMbFF} use experimental binding energies for $\Xi$ hyperon and extensive experimental data for $\Lambda$ hyperons, while ref \cite{Banik_2014_QOKgS} rely on effective field theory-based interactions calibrated with known hypernuclear data. For the present calculation, we use the above-mentioned 'DD2' family EoSs, which are available at CompOSE in tabulated forms \cite{obspmHome, doiCompOSEReference, Typel_2015_gH0QU}.

Further, the presence of the entire baryonic octet in the EoS model \cite{Marques_2017_dQf8a} is suggestive of a maximum mass which is higher than the maximum mass obtained in the presence of $\Lambda$ hyperons only. To include the effect of the entire baryonic octet, we use the EoS from ref \cite{Marques_2017_dQf8a} where $\sigma$-meson mediated hyperon interaction is considered. The EoS model is referred to as 'DD2Y', and the resulting maximum mass for cold NS sequence supports the recent observational limits \cite{Miller_2021_pAcGM, Riley_2021_kbhlr, Salmi_2022_wcauZ}. To model $\Delta$-admixed nuclear matter, the 'DD2Y' model is further extended to incorporate the $\Delta$ resonances by \cite{Raduta_2022_mFVlJ}. The authors in the original work took normalized meson-$\Delta$ baryon coupling constants (x$_{\sigma \Delta}$, x$_{\omega \Delta}$, x$_{\rho \Delta}$) to be (1.1, 1.1, 1.0), (1.2, 1.1, 1.0), (1.2, 1.3, 1.0), respectively. The EoSs are referred to as 'DD2Y$\Delta$(a)-(c)' corresponding to three $\Delta$ potential depths, $U_\Delta(\rho_{sat}) = $ -83, -124, -57 MeV, respectively \cite{Raduta_2022_mFVlJ}. For the present calculation, we use the mentioned tabulated EoSs available at CompOSE \cite{obspmHome, doiCompOSEReference, Typel_2015_gH0QU}.
    
It is worth mentioning the fact that the symmetry energy and slope of symmetry energy for the EoSs introduced above are consistent with the constraints drawn from PREX-II neutron skin thickness measurements \cite{Reinhard_2021_yDCxi, Essick_2021_AG8v4}. However, the EoSs exceed the smaller value of the symmetry energy slope parameter suitable for the CREX measurements \cite{Miyatsu_2023_XrwEA}. Further, the chosen set of EoSs, which covers a wide range of stiffnesses, not only places the nuclear incompressibility parameter within the acceptable range but also supports the negative symmetry incompressibilities favoured by astrophysical constraints \cite{Essick_2021_AG8v4, Essick_2021_fCXCe}. The models obtain maximum mass and corresponding radius values for static configurations within the ranges of 2.0 $M_{\odot}$ $\leq$ $M_{\text{max}}$ $\leq$ 2.4 $M_{\odot}$ and 12 $\lesssim$ R$_{1.4}$ $\lesssim$ 13.5 Km. Thus, the mass-radius constraints from NICER observations which suggest stiffer EoSs \cite{Cromartie_2019_NtRiY, Miller_2021_pAcGM, Riley_2019_oQUYh, Riley_2021_kbhlr, Salmi_2022_wcauZ} and the tidal deformability constraints from GW170817 \cite{Abbott_2017_TvRVX, Abbott_2018_EX9Nk, Landry_2019_HYTmg, Tews_2018_VRpfs}, which prefer softer EoSs, both are more or less satisfied with our chosen set of EoSs.

\subsection{Quark Matter}
\label{eos_quark}
In the cores of compact stars, the density of the matter exceeds symmetric nuclear matter saturation density significantly. Quantum chromodynamics (QCD) phase diagram suggests a phase transition from hadronic degrees of freedom to deconfined quark matter at such large densities \cite{Halasz_1998_ssGjO}. MIT bag model EOS is well adopted in the literature to reproduce the hadron-quark phase transition and study compact stars properties, core-collapse supernovae, etc., with a quark core \cite{Zdunik_2013_cA65e, _zel_2010_JsOjN, Fischer_2011_SOE3a}. The MIT Bag model provides a simple yet parameterizable framework incorporating confinement and asymptotic freedom, suitable for studying phase transitions and neutron star properties, albeit with limited treatment of quark interactions at large densities. In contrast, pQCD EoS offers a rigorous theoretical basis, scalability to high densities, and adherence to QCD principles. Lattice QCD studies of the thermodynamics of hot quark-gluon plasma demonstrate the importance of considering the interactions in the investigation of the phase structure of strongly interacting matter. Therefore, pQCD offers a more sophisticated framework than the MIT Bag model for constructing quark matter EoS at large densities where the perturbative approach is reliable. Considering this aspect, we adopt the EoS for interacting quark matter introduced in Ref \cite{Fraga_2014_1XoqE}, which incorporates the effects of interactions and includes inherent systematic uncertainties via dependence on the renormalization scale parameter. Thus, we employ an effective EoS for cold quark matter, which goes beyond the MIT bag model in certain aspects yet is equally straightforward. 

We utilize the Maxwell construction to describe the hadron-quark matter phase transition \cite{Glendenning_2001_zTYrc}. The pQCD EoS is so employed that the mass-energy density jump at phase transition is small so as to obtain stable and connected branches of HSs \cite{Alford_2013_7wL5W}. The mass-radius relations of resulting HSs obey the recent observational constraints well \cite{Miller_2021_pAcGM, Riley_2021_kbhlr, Salmi_2022_wcauZ} without violating the speed of sound conformal limit \cite{Annala_2020_6ap28}. The radius constraints provided by $\sim$ 1.4 M$_\odot$ PSR J0030+0451 are readily satisfied for the hybrid EoSs. However, a reduction in the maximum masses is observed for HSs. For our range of hybrid EoSs, we obtained 1.95 $M_{\odot}$ $\leq$ $M_{\text{max}}$ $\leq$ 2.2 $M_{\odot}$ to be the range for maximum masses.

\section{Description of Equilibrium Model}
\label{equilibrium_model}
The equilibrium configuration of compact objects within the context of general relativity is described in this section, assuming the gravitational constant $G = 1$ and speed of light $c = 1$. The relevant equations to calculate observables like mass, radius, deformability, moment of inertia, etc, are specified for non-rotating, slowly rotating and rapidly rotating stars. The energy-momentum tensor $T_{\mu \nu}$ within the star, assuming it to be a perfect fluid, is given by, 
\begin{equation}
  T_{\mu \nu} = \left( \varepsilon + P \right) u_{\mu} u_{\nu} + P g_{\mu \nu},
 \label{eqn1}   
\end{equation}
where, $\varepsilon$ is the energy density, $P$ is the fluid pressure, $u_{\mu}$ is the corresponding four velocity, and $g_{\mu \nu}$ is the Schwarzschild metric tensor.
 \subsection{Model for Non-rotating Stars}
 The metric tensor, which describes a non-rotating star, is spherically symmetric. In ($t$, $r$, $\theta$, $\varphi$) co-ordinate the metric is given by
 \begin{equation}
   ds^2 = - e^{2\nu(r)}dt^2 + e^{2\lambda(r)} dr^2 + r^2(d\theta^2 + \sin^2\theta d\varphi^2),
  \label{eqn2}
 \end{equation}
 where the $\nu(r)$ and $\lambda(r)$ are the metric functions to be determined from the equations of equilibrium, i.e., the Einstein equations. The mass-radius for non-rotating stars is finally obtained by integrating the Tolman-Oppenheimer-Volkoff (TOV) equations \cite{Tolman_1939_WVTIU, Oppenheimer_1939_oVOhJ},
 \begin{eqnarray}
     \frac{dM(r)}{dr} = 4\pi r^2 \varepsilon(r) \\
     \frac{dP(r)}{dr} = -\frac{(\varepsilon + P)(M + 4\pi r^3 P)}{r^2 (1 - \frac{2M}{r})}.
  \label{eqn3}
 \end{eqnarray}
 The equations are integrated from r $= 0$ to r $= R$, the radius of the star where the pressure P $= 0$.
 \subsection{Slowly Rotating Stars and Tidal Deformability}
 Following the Hartle-Thorne description for slowly rotating stars, we represent the metric tensor for the rotating configuration as \cite{Hartle_1967_7x5QB, Benhar_2005_gKaTj},
 \begin{eqnarray}
     & ds^2 = - e^{2\nu\left( r \right)}\left[ 1+2h\left( r, \theta \right) \right]dt^2 + e^{2\lambda(r)} \left[ 1 + \frac{2m\left( r, \theta \right)}{r-2M}\right] dr^2 \nonumber\\
     & + r^2 \left[ 1 + 2k\left( r, \theta \right) \right] \left[d\theta^2 + \sin^2\theta \left( d\varphi - \omega\left( r, \theta \right) dt \right)^2 \right]
  \label{eqn2a}
 \end{eqnarray}
 owing to the fact that the compact object retains symmetry with respect to $\varphi$. In the above expression, $\nu\left( r \right)$, $\lambda\left( r \right)$, and $M\left( r \right)$ are functions representing the background non-rotating configuration, while the perturbation terms $h\left( r, \theta \right)$, $m\left( r, \theta \right)$, and $k\left( r, \theta \right)$ describe the compact object rotating with an angular velocity $\Omega$. The term $\omega\left( r, \theta \right)$ represents the drag of the inertial frame, which has a maximum at the centre of the star and diminishes to zero far outside the star. Einstein's equations are expanded in the order of $\Omega$ to find the equilibrium configuration for a slowly rotating star. The lowest order $\Omega$ equations reproduce Eqs \ref{eqn3} and provide the metric for background non-rotating stars. It may be noted that the perturbation terms are functions of $r$ \& $\theta$. They are expanded with Legendre polynomials, $P_l\left( \cos \theta \right)$, keeping symmetry with respect to the $\theta = \pi/2$ plane. The equations of the next order in $\Omega$ for $l = 0$ are given by Eqs (97) and (100), and for $l = 2$ are given by Eqs (125) and (126) of ref. \cite{Hartle_1967_7x5QB}, and these equations are used to solve for perturbations concurrently with the background non-rotating equations. The utilization of the $t \varphi$ component of the Einstein equation in the context of determining the moment of inertia is given in Eqs (46)-(48) of ref. \cite{Hartle_1967_7x5QB}. The $l = 0$ perturbation equations are used to calculate the increase in the gravitational mass due to rotation, while the solution of $l = 2$ perturbation equations leads to the quadrupole moment of rotating stars. For this study, we consider perturbation terms of the order $>\mathcal{O}(\Omega^2)$ to calculate the moment of inertia, angular momentum, and quadrupole moment \cite{Benhar_2005_gKaTj, Yagi_2014_lZhJS}.

 Next, in the context of general relativity, we focus on stellar perturbations suitable for non-rotating or very slowly rotating stars. Using the Regge-Wheeler gauge for even parity perturbations, Einstein's equation leads to \cite{Thorne_1967_LSquP}
 \begin{eqnarray}
   H_0^{''} + \left( \frac{2}{ r} + \frac{\nu^{'} - \lambda^{'}}{2}\right) H_0^{'} + \left( \frac{2}{r^2} - \left[ 2+l\left(l+1\right) \right] \frac{e^\lambda}{r^2} \right. \nonumber\\
   \left. + \frac{9\lambda^{'} + 5\lambda^{'}}{2r} - \nu^{'2} + \frac{\nu^{'} + \lambda^{'}}{2rc_s^2} \right)H_0 = 0,  ~~
  \label{eqn4}
 \end{eqnarray}
 where, $H_0$ represents the perturbation in the metric, ${'} \equiv \frac{d}{dr}$ and $c_s^2 = \frac{dP}{d\varepsilon}$ is the speed of sound in the medium. Examining the r $=$ 0 behaviour of the star, we get the initial condition given by Eq (16) of Ref \cite{Hinderer_2008_ZsqiB}. The arbitrary constant of integration is fixed by making the solution continuous across the surface (at r $=$ R). Finally, the tidal love number $k_2$ can be expressed as
 \begin{eqnarray}
      & k_2 = \frac{8C^5}{5} \left(1 - 2C\right)^2 \left[2 + 2C\left(y - 1\right) - y\right] \nonumber\\ 
      & \times \left\lbrace 2C \left[6 - 3y + 3C(5y - 8) \right] \right. \nonumber\\
      & + 4C^3 \left[13 - 11y + C(3y - 2) + 2C^2 (1 + y)\right] \nonumber\\
      & \left. + 3(1 - 2C)^2 \left[2 - y + 2C(y - 1)\right] \ln(1 - 2C)\right\rbrace^{-1}
    \label{eqn5}
 \end{eqnarray}
 where, the quantity $C = M/R$ is compactness of the star and $y = {R H_0^{'}\left( R \right)}/{H_0(R)}$ depends on the solution at surface. In the definition of compactness $C$, $M$ is the total gravitating mass of the rotating body, while $R$ is the equatorial radius $R_{eq}$ of the star. For calculation purpose we define the dimensionless tidal deformability $\overline{\Lambda}$ to be
 \begin{equation}
   \overline{\Lambda} = \frac{2}{3} k_2 C^{-5}
     \label{eqn6}
 \end{equation}
 \subsection{Model for Rapidly Rotating Stars}
 Extension of the Newtonian self-consistent field method to the general relativistic scenario enables the study of rapidly rotating relativistic stars \cite{Komatsu_1989_hBLFz}. Then, assuming the azimuthal symmetry and reflection symmetry about the equatorial plane, the metric tensor of the space-time can be represented as
 \begin{eqnarray}
     ds^2 = & - e^{ \gamma + \rho } dt^2 + e^{2\alpha} \left( dr^2 + r^2 d\theta^2 \right) \nonumber\\
     & + e^{ \gamma - \rho } r^2 \sin^2 \theta \left( d\varphi - \omega dt \right)^2 ,~~ 
  \label{eqn2b}
 \end{eqnarray}
 where the metric potentials $\gamma$, $\rho$, $\alpha$, and $\omega$ are functions of $r$ and $\theta$. Furthermore, we employ the numerical technique of RNS code to calculate the mass, radius, moment of inertia, and quadrupole moment of rapidly rotating NSs \cite{Koranda_1997_82RG7, Laarakkers_1999_PPeZV}.

 To describe different rotating sequences of NSs, we define the dimensionless spin parameter to be,
 \begin{equation}
  \chi = \frac{J}{M^2}
     \label{eqn7}
 \end{equation}
 where $J$ is the angular momentum of the rotating star. It must be noted that $M$, in the whole description, represents the total gravitating mass of the body.

 We define a few useful dimensionless quantities like the moment of inertia $\overline{I}$, quadrupole moment $\overline{Q}$ as follows,
 \begin{eqnarray}
  \overline{I} &=& \frac{I}{M^3} \nonumber\\
  \overline{Q} &=& \frac{Q}{M^3\chi^2}
     \label{eqn8}
 \end{eqnarray}

\twocolumngrid
\section{Results and Discussion}
\label{result_discussion}
In this section, we summarize the universal relations among NS higher multipole moments, such as the moment of inertia and quadrupole moment for both slowly and rapidly rotating NSs and extend the relations for connected stable branches of HSs. We present our results in comparison to existing quasi-universal relations \cite{Li_2023_wnUwG, Raduta_2020_YKkHx, Lenka_2019_VAoni, Marques_2017_dQf8a} to clarify the role of exotic degrees of freedom in determining the relations. The possibility of hadron-quark phase transition is subsequently considered at sufficiently high densities. In the description of Maxwell's construction of phase transition, we have obtained stable branches of HSs for all of our hadron EoSs. The properties of the HSs are investigated and found to be consistent with current astrophysical constraints on mass-radius. The inferences drawn on radii, tidal deformabilities, etc. of NSs from events like GW170817 \cite{Tews_2018_VRpfs, Paschalidis_2018_AetFT, Most_2018_JHlGX, Malik_2018_qAfAT, Essick_2020_enQBF, De_2018_LcNIX} further justify our EoSs. After that, the universal relations for these HSs are investigated and compared with already available literature. Thus, we have explored universal relations for compact objects with almost all the exotic degrees of freedom.

Figures (start) \ref{fig1_eos}(a) and \ref{fig1_eos}(b) show the hadron EoSs and the hybrid EoSs and the corresponding speed of sound squared $c_s^2$, respectively. In Figure \ref{fig1_eos}, the solid circles, solid squares, triangles, and solid triangles represent the EoS points which lead to 1.4 $M_\odot$ configuration, hadron-quark phase transition, maximum mass configuration with hadronic core, and maximum mass configuration with quark core, respectively. Figure \ref{fig1_eos}(a) shows that for both hadronic and hybrid EoSs, pressure increases with energy density or baryon density. However, the phase transition leads to strongly interacting quark matter and a considerable decrease in pressure compared to hadronic EoSs at a given density. The position of different mass points and phase transition points on the EoSs indicates the presence of a sizable quark-matter core in massive NSs \cite{Annala_2020_6ap28}. This can further be related to the behaviour of the speed of sound, as depicted in Figure \ref{fig1_eos}(b). The $c_s^2$ is seen to be increasing monotonically with $\varepsilon$ for purely nucleonic EoS DD2. However, the appearance of exotic degrees of freedom leads to a change in the trend of $c_s^2$. Firstly, the onset of hyperons and $\Delta$-baryon reduces the speed of sound significantly. The abrupt changes in $c_s^2$ occur in the lower density region, indicating a region of solid-liquid type phase transition. The speed of sound further drops on the appearance of deconfined quark degrees of freedom. The hadron-quark phase transition is accompanied by another abrupt change in $c_s^2$, eventually leading to a phase where $c_s^2$ remains almost constant, the value of which stays within the conformal bound on $c_s^2$.

Figure \ref{fig2_M-R} shows mass-radius relations for NSs with hadronic cores with exotic degrees of freedom and HSs with quark matter cores. Figure \ref{fig2_M-R}(a) depicts the mass-radius relation for static configurations along with the constraints set by current astrophysical observations. It is observed that the maximum mass configurations for our choice of EoSs satisfy 1.95 $M_{\odot}$ $\leq$ $M_{\text{max}}$ $\leq$ 2.4 $M_{\odot}$. The radius for 1.4 $M_{\odot}$, on the other hand, of  12 $\lesssim$ R$_{1.4}$ $\lesssim$ 13.5 Km. Thus, our EoSs predict NS models that are consistent with the NICER inferences of pulsars PSR J0030+0451 and PSR J0740+6620 \cite{Miller_2019_dFNjM, Miller_2021_pAcGM, Riley_2019_oQUYh, Riley_2021_kbhlr}. The mass-equatorial radius relation of maximally rotating configurations, referred to as the 'Keplerian' sequence, is shown in Figure \ref{fig2_M-R}(b). This shows an increase in both mass and equatorial radius to such an extent that it exceeds the upper limit set for massive pulsar PSR J0740+6620 for all the EoSs. It is worth mentioning the fact that the GW190814's secondary object mass remains in the lower mass gap of 2.5-5 $M_{\odot}$ \cite{Bailyn_1998_NQC7Q, _zel_2010_2eN7k, _zel_2012_kDmbl} between observed NSs and BHs. It is heavier than both the massive millisecond pulsar PSR J0740+6620 \cite{Fonseca_2021_SOyMc}, and the 1.61-2.52 $M_{\odot}$ primary component of GW190425 \cite{Abbott_2020_WX55j}. The GW170817 merger remnant of $\sim$ 2.7 $M_{\odot}$ BH \cite{Abbott_2019_ION0q} and 2.74$\pm$0.21 $M_{\odot}$ (68\% confidence) millisecond pulsar PSR J1748-2021B \cite{Freire_2008_smnqk} both are comparable to the mass of GW190814's secondary. Therefore, the nature of the GW190814's secondary object remains inconclusive. However, the obtained mass range 2.50–2.67 $M_{\odot}$ for GW190814's secondary compact object \cite{Abbott_2020_jIBsD} is attainable with maximally rotating configurations for the chosen set of EoSs.

\onecolumngrid
\onecolumngrid
\begin{table}[t]
\centering
\captionsetup{justification=raggedright}
\caption{Parameters of fitting of C-$\overline{\Lambda}$ relation using Eq. (\ref{eqn9}). Reduced $\chi^2$ parameter of fitting is presented in the last column $\left( \chi_{\text{red}}^2 \right)$.}
\label{table1}
\begin{tabularx}{\textwidth}{|p{0.4cm} p{0.5cm} X X X X X X X|}
\hline
\textbf{y} & \textbf{x} & \textbf{Comp} & $\boldsymbol{~~~~~a_0}$ & $\boldsymbol{~~~~~a_1}$ & $\boldsymbol{~~~~~a_2}$ & $\boldsymbol{~~~~~a_3}$ & $\boldsymbol{~~~~~a_4}$ & $\boldsymbol{~~~~~\chi_{\text{red}}^2}$ \\ 
\hline\hline
~ & ~ & N, Y, $\Delta$ & $3.5868\times10^{-1}$ & $-3.6527\times10^{-2}$ & $8.5581\times10^{-4}$ & ~ & ~ & $3.9635\times10^{-6}$ \\
$C$ & $\overline{\Lambda}$ & N, Y, $\Delta$, q & $3.5893\times10^{-1}$ & $-3.6593\times10^{-2}$ & $8.5970\times10^{-4}$ & ~ & ~ & $3.5397\times10^{-6}$ \\
~ & ~ & N, Y, $\Delta$ & $3.6127\times10^{-1}$ & $-4.0917\times10^{-2}$ & $2.7693\times10^{-3}$ & $-2.9642\times10^{-4}$ & $1.4749\times10^{-5}$ & $3.3282\times10^{-6}$ \\
~ & ~ & N, Y, $\Delta$, q & $3.5997\times10^{-1}$ & $-4.0140\times10^{-2}$ & $2.6090\times10^{-3}$ & $-2.8277\times10^{-4}$ & $1.4343\times10^{-5}$ & $2.9072\times10^{-6}$ \\
\hline
\end{tabularx}
\end{table}
\onecolumngrid
\begin{table}
\centering
\captionsetup{justification=raggedright}
\caption{Parameters of C-$\overline{\Lambda}$, $\overline{I}$-$\overline{\Lambda}$, $\overline{Q}$-$\overline{\Lambda}$, $\overline{I}$-$\overline{Q}$ universal relations using Eq. (\ref{eqn10}). Reduced $\chi^2$ parameter of fitting is presented in the last column $\left( \chi_{\text{red}}^2 \right)$.}
\label{table2}
\begin{tabularx}{\textwidth}{|p{0.4cm} p{0.8cm} X X X X X X X|}
\hline
\textbf{y} & \textbf{x} & \textbf{Comp} &~~ $\boldsymbol{a_0}$ & ~~~$\boldsymbol{a_1}$ & ~~~~$\boldsymbol{a_2}$ & ~~~~ $\boldsymbol{a_3}$ & ~~~~ $\boldsymbol{a_4}$ & ~~~~ $\boldsymbol{\chi_{\text{red}}^2}$ \\ 
\hline\hline
~ & ~ & N, Y, $\Delta$ & $1.3941 \times 10^{0}$ & $1.2053 \times 10^{-1}$ & $1.1174 \times 10^{-2}$ & ~ & ~ & $4.8933 \times 10^{-5}$ \\
$\overline{I}$ & $\overline{\Lambda}$ & N, Y, $\Delta$, q & $1.3828 \times 10^{0}$ & $1.2388 \times 10^{-1}$ & $1.0947 \times 10^{-2}$ & ~ & ~ & $4.0534 \times 10^{-5}$ \\
~ & ~ & N, Y, $\Delta$ & $1.5053 \times 10^{0}$ & $5.1481 \times 10^{-2}$ & $2.4523 \times 10^{-2}$ & $-9.1879 \times 10^{-4}$ & $1.6269 \times 10^{-5}$ & $1.1656 \times 10^{-6}$ \\
~ & ~ & N, Y, $\Delta$, q & $1.5030 \times 10^{0}$ & $5.3254 \times 10^{-2}$ & $2.4077 \times 10^{-2}$ & $-8.7307 \times 10^{-4}$ & $1.4628 \times 10^{-5}$ & $9.8163 \times 10^{-7}$ \\
\hline
~ & ~ & N, Y, $\Delta$ & $-1.3091 \times 10^{-1}$ & $3.1287 \times 10^{-1}$ & $-1.0453 \times 10^{-3}$ & ~ & ~ & $2.1305 \times 10^{-4}$ \\
$\overline{Q}$ & $\overline{\Lambda}$ & N, Y, $\Delta$, q & $-1.5730 \times 10^{-1}$ & $3.2068 \times 10^{-1}$ & $-1.5708 \times 10^{-3}$ & ~ & ~ & $1.7174 \times 10^{-4}$ \\
~ & ~ & N, Y, $\Delta$ & $1.8090 \times 10^{-1}$ & $1.0282 \times 10^{-1}$ & $4.5331 \times 10^{-2}$ & $-4.0277 \times 10^{-3}$ & $1.1744 \times 10^{-4}$ & $3.2774 \times 10^{-6}$ \\
~ & ~ & N, Y, $\Delta$, q & $1.8304 \times 10^{-1}$ & $1.010 \times 10^{-1}$ & $4.5812 \times 10^{-2}$ & $-4.0785 \times 10^{-3}$ & $1.1930 \times 10^{-4}$ & $2.9837 \times 10^{-6}$ \\
\hline
~ & ~ & N, Y, $\Delta$ & $1.4732 \times 10^{0}$ & $3.7114 \times 10^{-1}$ & $1.4010 \times 10^{-2}$ & ~ & ~ & $3.9775 \times 10^{-5}$ \\
$\overline{I}$ & $\overline{Q}$ & N, Y, $\Delta$, q & $1.4819 \times 10^{0}$ & $3.6217 \times 10^{-1}$ & $1.4220 \times 10^{-1}$ & ~ & ~ & $3.5297 \times 10^{-5}$ \\
~ & ~ & N, Y, $\Delta$ & $1.3995 \times 10^{0}$ & $5.2866 \times 10^{-1}$ & $4.1204 \times 10^{-2}$ & $1.9237 \times 10^{-2}$ & $-2.7317 \times 10^{-4}$ & $5.2622 \times 10^{-6}$ \\
~ & ~ & N, Y, $\Delta$, q & $1.3944 \times 10^{0}$ & $5.4260 \times 10^{-1}$ & $2.860 \times 10^{-2}$ & $2.3849 \times 10^{-2}$ & $-8.6344 \times 10^{-4}$ & $4.6436 \times 10^{-6}$ \\
\hline
\end{tabularx}
\end{table}
\onecolumngrid
\begin{table}[t]
\centering
\captionsetup{justification=raggedright}
\caption{Parameters of $\overline{I}$-C, $\overline{Q}$-C relations using Eq. (\ref{eqn11}) and $\overline{I}$-$\overline{Q}$ relation using Eq. (\ref{eqn10}). Reduced $\chi^2$ of fitting is presented in the last column $\left( \chi_{\text{red}}^2 \right)$.}
\label{table3}
\begin{tabularx}{\textwidth}{|p{0.2cm} p{0.8cm} p{1.8cm} p{1.2cm} X X X X X X|}
\hline

\textbf{y} & ~~ \textbf{x} & \textbf{Comp} & $~\boldsymbol{\chi}$ &~~ $\boldsymbol{a_0}$ & ~~~$\boldsymbol{a_1}$ & ~~~~$\boldsymbol{a_2}$ & ~~~~ $\boldsymbol{a_3}$ & ~~~~ $\boldsymbol{a_4}$ & ~~~~ $\boldsymbol{\chi_{\text{red}}^2}$ \\
\hline\hline

 ~ & ~ & N, Y, $\Delta$ & slow &  & 5.4899$\times 10 ^{-1}$ & 3.1766$\times 10 ^{-1}$ & -9.7468$\times 10 ^{-3}$ & 1.5699$\times 10 ^{-4}$ & 3.5187$\times 10 ^{-2}$ \\
~ & ~ & N, Y, $\Delta$, q & &  & 5.3480$\times 10 ^{-1}$ & 3.2232$\times 10 ^{-1}$ & -1.0217$\times 10 ^{-2}$ & 1.7170$\times 10 ^{-4}$ & 3.5168$\times 10 ^{-2}$ \\

~ & ~ & N, Y, $\Delta$ & 0.2 &  & 5.3382$\times 10 ^{-1}$ & 3.2351$\times 10 ^{-1}$ & -1.0736$\times 10 ^{-2}$ & 1.9324$\times 10 ^{-4}$ & 3.8629$\times 10 ^{-2}$ \\
~ & ~ & N, Y, $\Delta$, q & &  & 5.1108$\times 10 ^{-1}$ & 3.3162$\times 10 ^{-1}$ & -1.1612$\times 10 ^{-2}$ & 2.2339$\times 10 ^{-4}$ & 3.7650$\times 10 ^{-2}$ \\

~ & ~ & N, Y, $\Delta$ & 0.4 &  & 5.1559$\times 10 ^{-1}$ & 3.2680$\times 10 ^{-1}$ & -1.1850$\times 10 ^{-2}$ & 2.2556$\times 10 ^{-4}$ & 4.5353$\times 10 ^{-2}$ \\
$\overline{I}$ & ~~ $C$ & N, Y, $\Delta$, q & &  & 4.9521$\times 10 ^{-1}$ & 3.3405$\times 10 ^{-1}$ & -1.2627$\times 10 ^{-2}$ & 2.5227$\times 10 ^{-4}$ & 4.4304$\times 10 ^{-2}$ \\

~ & ~ & N, Y, $\Delta$ & 0.6 &  & 4.3975$\times 10 ^{-1}$ & 3.3469$\times 10 ^{-1}$ & -1.4432$\times 10 ^{-2}$ & 2.9407$\times 10 ^{-4}$ & 9.2824$\times 10 ^{-2}$ \\
~ & ~ & N, Y, $\Delta$, q & &  & 4.2043$\times 10 ^{-1}$ & 3.4158$\times 10 ^{-1}$ & -1.5157$\times 10 ^{-2}$ & 3.1856$\times 10 ^{-4}$ & 9.0804$\times 10 ^{-2}$ \\

~ & ~ & N, Y, $\Delta$ & Kep. &  & 5.5124$\times 10 ^{-1}$ & 2.410$\times 10 ^{-1}$ & -9.6833$\times 10 ^{-3}$ & 1.8739$\times 10 ^{-4}$ & 4.7559$\times 10 ^{-2}$ \\
~ & ~ & N, Y, $\Delta$, q & &  & 5.4436$\times 10 ^{-1}$ & 2.4318$\times 10 ^{-1}$ & -9.8754$\times 10 ^{-3}$ & 1.9287$\times 10 ^{-4}$ & 4.6633$\times 10 ^{-2}$ \\

\hline
~ & ~ & N, Y, $\Delta$ & slow &  & -5.7681$\times 10 ^{-1}$ & 3.8503$\times 10 ^{-1}$ & -2.5088$\times 10 ^{-2}$ & 5.7674$\times 10 ^{-4}$ & 6.5196$\times 10 ^{-3}$ \\
~ & ~ & N, Y, $\Delta$, q & &  & -5.8033$\times 10 ^{-1}$ & 3.8615$\times 10 ^{-1}$ & -2.5198$\times 10 ^{-2}$ & 5.8010$\times 10 ^{-4}$ & 6.5116$\times 10 ^{-3}$ \\

~ & ~ & N, Y, $\Delta$ & 0.2 &  & -7.7819 $\times 10 ^{-1}$ & 4.5341$\times 10 ^{-1}$ & -3.3422$\times 10 ^{-2}$ & 8.8988$\times 10 ^{-4}$ & 2.6436$\times 10 ^{-2}$ \\
~ & ~ & N, Y, $\Delta$, q & &  & -7.8556$\times 10 ^{-1}$ & 4.5571$\times 10 ^{-1}$ & -3.3631$\times 10 ^{-2}$ & 8.9574$\times 10 ^{-4}$ & 2.7366$\times 10 ^{-2}$ \\

~ & ~ & N, Y, $\Delta$ & 0.4 &  & -5.0847$\times 10 ^{-1}$ & 3.2557$\times 10 ^{-1}$ & -2.1005$\times 10 ^{-2}$ & 4.9395$\times 10 ^{-4}$ & 6.2535$\times 10 ^{-3}$ \\
$\overline{Q}$ & ~~$C$ & N, Y, $\Delta$, q & &  & -5.1634$\times 10 ^{-1}$ & 3.2845$\times 10 ^{-1}$ & -2.1322$\times 10 ^{-2}$ & 5.0508$\times 10 ^{-4}$ & 6.1015$\times 10 ^{-3}$ \\

~ & ~ & N, Y, $\Delta$ & 0.6 &  & -3.0857$\times 10 ^{-1}$ & 2.2844$\times 10 ^{-1}$ & -1.3507$\times 10 ^{-2}$ & 2.9378$\times 10 ^{-4}$ & 9.9778$\times 10 ^{-3}$ \\
~ & ~ & N, Y, $\Delta$, q & &  & -3.1913$\times 10 ^{-1}$ & 2.3212$\times 10 ^{-1}$ & -1.3891$\times 10 ^{-2}$ & 3.0649$\times 10 ^{-4}$ & 9.7031$\times 10 ^{-3}$ \\

~ & ~ & N, Y, $\Delta$ & Kep. &  & -9.9268$\times 10 ^{-2}$ & 1.3058$\times 10 ^{-1}$ & -6.2730$\times 10 ^{-3}$ & 1.150$\times 10 ^{-4}$ & 7.8989$\times 10 ^{-4}$ \\
~ & ~ & N, Y, $\Delta$, q & &  & -1.0168$\times 10 ^{-1}$ & 1.3133$\times 10 ^{-1}$ & -6.3410$\times 10 ^{-3}$ & 1.1694$\times 10 ^{-4}$ & 7.1575$\times 10 ^{-4}$ \\

\hline
 ~ & ~ & N, Y, $\Delta$ & slow & 1.3995$\times 10 ^{0}$ & 5.2866$\times 10 ^{-1}$ & 4.1204$\times 10 ^{-2}$ & 1.9237$\times 10 ^{-2}$ & -2.7317$\times 10 ^{-4}$ & 5.2622$\times 10 ^{-6}$ \\
~ & ~ & N, Y, $\Delta$, q & & 1.3944$\times 10 ^{0}$ & 5.4260$\times 10 ^{-1}$ & 2.860$\times 10 ^{-2}$ & 2.3849$\times 10 ^{-2}$ & -8.6344$\times 10 ^{-4}$ & 4.6436$\times 10 ^{-6}$ \\

~ & ~ & N, Y, $\Delta$ & 0.2 & 1.4443$\times 10 ^{0}$ & 6.8393$\times 10 ^{-1}$ & -2.4977$\times 10 ^{-1}$ & 1.7121$\times 10 ^{-1}$ & -2.4810$\times 10 ^{-2}$ & 1.5393$\times 10 ^{-4}$ \\
~ & ~ & N, Y, $\Delta$, q & & 1.4390$\times 10 ^{0}$ & 7.1154$\times 10 ^{-1}$ & -2.8157$\times 10 ^{-1}$ & 1.8449$\times 10 ^{-1}$ & -2.6660$\times 10 ^{-2}$ & 1.6325$\times 10 ^{-4}$ \\

~ & ~ & N, Y, $\Delta$ & 0.4 & 1.5062$\times 10 ^{0}$ & 4.9936$\times 10 ^{-1}$ & 4.9936$\times 10 ^{-1}$ & -5.8355$\times 10 ^{-3}$ & 2.7121$\times 10 ^{-3}$ & 5.2904$\times 10 ^{-6}$ \\
$\overline{I}$ & ~~ $\overline{Q}$ & N, Y, $\Delta$, q & & 1.5047$\times 10 ^{0}$ & 5.0499$\times 10 ^{-1}$ & 9.6454$\times 10 ^{-2}$ & -3.0447$\times 10 ^{-3}$ & 2.2931$\times 10 ^{-3}$ & 4.7362$\times 10 ^{-6}$ \\

~ & ~ & N, Y, $\Delta$ & 0.6 & 1.5038$\times 10 ^{0}$ & 6.5099$\times 10 ^{-1}$ & 4.5158$\times 10 ^{-2}$ & 6.8310$\times 10 ^{-3}$ & 1.1563$\times 10 ^{-3}$ & 5.2472$\times 10 ^{-6}$ \\
~ & ~ & N, Y, $\Delta$, q & & 1.5021$\times 10 ^{0}$ & 6.5863$\times 10 ^{-1}$ & 3.5559$\times 10 ^{-2}$ & 1.1398$\times 10 ^{-2}$ & 4.2440$\times 10 ^{-4}$ & 4.5516$\times 10 ^{-6}$ \\

~ & ~ & N, Y, $\Delta$ & Kep. & 1.5149$\times 10 ^{0}$ & 6.4862$\times 10 ^{-1}$ & 1.3484$\times 10 ^{-1}$ & -4.4522$\times 10 ^{-2}$ & 9.0956$\times 10 ^{-3}$ & 7.3224$\times 10 ^{-5}$ \\
~ & ~ & N, Y, $\Delta$, q & & 1.5121$\times 10 ^{0}$ & 6.6331$\times 10 ^{-1}$ & 1.1498$\times 10 ^{-1}$ & -3.4608$\times 10 ^{-2}$ & 7.4554$\times 10 ^{-3}$ & 7.2438$\times 10 ^{-5}$ \\
\hline
\end{tabularx}
\end{table}
\twocolumngrid
\subsection{Relations for slowly rotating stars}
\label{RD_slowNS}
The form of the Love number $k_2$ is given by Eq. \ref{eqn5}, whereas tidal deformability $\overline{\Lambda}$ of NS is given in Eq. \ref{eqn6}. The numerical results of $k_2$ are shown in Fig. \ref{fig3_k2-C}. It can be seen from this figure that $k_2$ decreases monotonically with $C$ for $C \geq $ 0.1. However, the trend changes around $C \lesssim$ 0.1, which is less important from an NS phenomenology perspective. Noting from Eq. \ref{eqn6} that $\overline{\Lambda}$ scales as $\overline{\Lambda} \propto k_2 C^{-5}$ with compactness $C$ and $k_2$ decreases monotonically with $C$ we can expect a universal $\overline{\Lambda} \propto C^{-6}$ behavior. Fig. \ref{fig4_C-L}, where $\overline{\Lambda}$ is presented in logarithmic scale, visually confirms this expected universal behaviour between C and $\overline{\Lambda}$. The relation between C and $\overline{\Lambda}$ is further investigated using a scaling formula,
\begin{equation}
  C = \sum\limits_{n=0}^{m} a_n \left( \ln \overline{\Lambda} \right)^n .
 \label{eqn9}
\end{equation}
This relation is explored in refs \cite{Maselli_2013_KEJBr, Yagi_2017_uWwLw, Raduta_2020_YKkHx, Yeung_2021_dc0JE, Burikham_2022_Kq9sd, Carlson_2023_0TLyw, Li_2023_wnUwG} and several other papers, which consider nucleonic, hadronic, hot nuclear matter, hybrid EoS models. We check the universality of C-$\overline{\Lambda}$ relation at zero temperature using Eq \ref{eqn9} for $m=2$ and $m=4$. We find universal relations separately for the cases of (a) hadronic EoSs only (N, Y, $\Delta$ composition) and (b) hadronic EoSs + hybrid EoSs (N, Y, $\Delta$, q composition). We compare our $m=2$ relation determined for case (a) with the one developed in ref \cite{Maselli_2013_KEJBr}. We find an overall agreement with Ref \cite{Maselli_2013_KEJBr} where the relation is obtained using APR4, MS1, and H4 EoSs. The C-$\overline{\Lambda}$ relation from our calculation, however, predicts a slightly smaller value of tidal deformability $\overline{\Lambda}$ compared to ref \cite{Maselli_2013_KEJBr}. The result is consistent with earlier findings on properties of compact stars with exotic degrees of freedom \cite{Li_2019_BXUgg, Li_2020_il2wz}. The relation is also provided in refs \cite{Yagi_2017_uWwLw, Li_2023_wnUwG, Raduta_2022_mFVlJ} with $m=2$. The role of temperature and $\beta$-equilibrium on the relation considering hadronic EoSs with few exotic degrees of freedom is discussed in ref \cite{Raduta_2022_mFVlJ} while ref \cite{Li_2023_wnUwG} provided the relation for compact objects with heavy baryons at zero temperature and $\beta$-equilibrium. Comparison of the $m=2$ and $m=4$ results with those from ref \cite{Li_2023_wnUwG} reveals an excellent match. Then, we have extended the relation for case (b) with both $m=2$ and $m=4$. In Fig. \ref{fig4_C-L}, only the fitting function using $m=4$ in Eq. \ref{eqn9} for case (b) is shown in the top panel, while the bottom panel presents the corresponding fractional error between the computed results and the universal C-$\overline{\Lambda}$ function. The deviation of computed results from universal relation is found to be at the level of a few per cent only. The coefficients of the relations for both cases (a) and (b) are given in Table \ref{table1} for both $m=2$ and $m=4$ along with $\chi_{\text{red}}^2$ of the fitting. It is noted that the appearance of a quark core tends to further decrease the deformability of the stars (Fig. \ref{fig3_k2-C}). Consequently, the C-$\overline{\Lambda}$ relation is also slightly affected due to the presence of deconfined quarks inside the core. However, the relative error of our C-$\overline{\Lambda}$ relation remains $\lesssim$ 4\%. Results from our calculation are in overall agreement with the C-$\overline{\Lambda}$ relations obtained in refs \cite{Godzieba_2021_UBDLe, Suleiman_2024_7cKmd}, where authors used a very large number of EOSs. The relative error for the relation in ref \cite{Suleiman_2024_7cKmd} using an EoS set constructed with the Gaussian process and constrained from current astrophysical observations is higher than our results. It is important to note that the parameters of our relations exhibit larger asymptotic standard errors for HSs compared to those for NSs. Ref \cite{Legred_2024_WRfSP} also obtained larger relative errors for C-$\overline{\Lambda}$ relation while introducing phase transition phenomenology.

Next, we investigate the universal relations associated with the I-Love-Q trio for our collection of EoSs and generalize them for stable HSs, considering slowly rotating stars. We explore relations between $\overline{I}$, $\overline{\Lambda}$, and $\overline{Q}$ employing the method introduced in Refs \cite{Yagi_2013_cPzvP, Yagi_2013_yxUgU},
\begin{equation}
 \ln y = \sum\limits_{n=0}^{m} a_n \left( \ln x \right)^n
 \label{eqn10}
\end{equation}
where the equation is used for ($\overline{\Lambda}$, $\overline{I}$), ($\overline{\Lambda}$, $\overline{Q}$), and ($\overline{Q}$, $\overline{I}$) pairs of variables treated as (x, y). The quadrupole moment and moment of inertia for slowly rotating stars are calculated using the perturbative approach of Ref \cite{Benhar_2005_gKaTj} where the spin parameter $\chi$ is taken to be a constant (10$^{-3}$). Using Eq. \ref{eqn10} and taking $m=2$ and $m=4$, we determine all these relations for both compositions (a) and (b) and compare with existing literature \cite{Baub_ck_2013_gJPHq, Suleiman_2024_7cKmd, Legred_2024_WRfSP, Urbanec_2013_wNmiD, Yeung_2021_dc0JE}. We summarize the fitting parameters along with $\chi_{\text{red}}^2$ in Table \ref{table2}. Coefficients of the I-Love-Q relations with composition (a) and $m=4$ are in overall agreement with those provided in table IV of the ref \cite{Wang_2022_1ft2d}, where the authors modelled NSs employing the relativistic Brueckner-Hartree-Fock theory in the full Dirac space and obtained the relations using the same method as refs \cite{Yagi_2013_cPzvP, Yagi_2013_yxUgU}. Further, the $m=4$ parameters of I-Love-Q relations obtained for NS EoSs with heavy baryon degrees of freedom for spin parameter $\chi \ll 1$ in Ref \cite{Li_2023_wnUwG} are in good agreement with our obtained results for composition (a). Then, we show the computed results in Figs. \ref{fig5_I-L_sr}-\ref{fig6_I-Q_sr} for all these pairs of variables with $m=4$. Similar to Fig. \ref{fig4_C-L}, computed results for EoSs with case (b) are presented in the upper panels of these figures. Corresponding fractional differences between the functions of universal relation and numerical results are shown in the bottom panels of the mentioned figures. The obtained results more or less agree with the earlier relations for HSs treated within the slow-rotation approximation \cite{Yeung_2021_dc0JE, Burikham_2022_Kq9sd}. Although the relative errors in our calculation do not deviate much for HS models, we observe a general trend of having larger asymptotic standard errors for fitting parameters of HS relations. The argument implies that the inclusion of additional degrees of freedom in the core composition adds further to the variability of the obtained relations.

Due to the universality observed in the C-$\overline{\Lambda}$ relation in present work and several other works in literature \cite{Maselli_2013_KEJBr, Godzieba_2021_UBDLe, Suleiman_2024_7cKmd, Legred_2024_WRfSP} and the $\overline{I}$-$\overline{\Lambda}$-$\overline{Q}$ relation, it is plausible to propose universality in $\overline{I}$-C-$\overline{Q}$ trio. We attempt exploring the universal relations employing the function suggested in Ref \cite{Yagi_2013_yxUgU}:

\begin{equation}
  y = \sum\limits_{n=1}^{4} a_n \left( \text{C}^{-1} \right)^n .
 \label{eqn11}
\end{equation}
Assuming y $= \overline{I}$ or $\overline{Q}$ we investigate $\overline{I}$-C and $\overline{Q}$-C relations. Fig \ref{fig8_I-C} (top) with legend $\chi = $ 10$^{-3}$ shows the EoS insensitive trend followed between $\overline{I}$ and C for slowly rotating compact objects. On the other hand, Fig \ref{fig9_Q-C} (top) with legend $\chi = $ 10$^{-3}$ shows numerical results for $\overline{Q}$-C pair for the mentioned fixed spin. In Fig \ref{fig10_I-Q} (top) computed results for $\overline{I}$-$\overline{Q}$ pair with spin parameter $\chi = 10^{-3}$ is shown. Again, the parameters for $\overline{I}$-C and $\overline{Q}$-C relations using eq. \ref{eqn11} and for $\overline{I}$-$\overline{Q}$ relation using eq. \ref{eqn10} with $m = 4$ are estimated considering both the compositions (a) and (b). The final results are summarized in table \ref{table3} as 'slow'. The functions of universal relations in all these figures are presented for composition (b) only, as earlier. The lower panels of Figs. \ref{fig8_I-C}-\ref{fig10_I-Q} displays relative errors between the fit functions and computed results. It is important to note that for the $\overline{I}$-C and $\overline{Q}$-C relations, relative errors are relatively higher in comparison to the $\overline{I}$-$\overline{\Lambda}$ and $\overline{Q}$-$\overline{\Lambda}$. Ref \cite{Breu_2016_WqIRD} derived $\overline{I}$-C relation for slowly rotating NSs using a set of nuclear EoSs following the method of ref \cite{Lattimer_2005_z0KLn} to obtain relative difference of $\sim$ 9\%. Our $\overline{I}$-C relation for slowly rotating stars with core composition (a) obtains a relatively lower fractional difference. Comparison with the relations of refs \cite{Baub_ck_2013_gJPHq, Li_2023_wnUwG} for slowly rotating stars implies reasonable agreement with our result for composition (a). The coefficients of $\overline{I}$-C-$\overline{Q}$ relations obtained for HSs have higher asymptotic standard errors compared to the coefficients for composition (a).

\subsection {Relations for rapidly rotating stars}
\label{RD_rapidNS}
Based on the discussion of universal relations for slowly rotating stars, it is now plausible to examine the validity of $\overline{I}$-C-$\overline{Q}$ relations for rapidly rotating stars. The universality of the $\overline{I}$-$\overline{Q}$ relation breaks down in rapidly rotating stars \cite{Doneva_2013_8E7jv}. However, one can obtain the universal relations for rapidly rotating stars along sequences of constant normalized spin parameters, such as $\chi$ \cite{Doneva_2014_K2OBz}. 

First, in fig \ref{fig2b_chi}, we present the $\chi_{\text{max}}$ value, which corresponds to the Keplerian sequence, as a function of compactness (top) and gravitational mass (bottom), respectively. We have found that initially $\chi_{\text{max}}$ increases with the increasing central density of the compact object followed by a saturation near C $\sim$ 0.05, which is equivalent to gravitational mass M $\lesssim$ 1.0 M$_{\odot}$. At saturation $\chi_{\text{max}}$ lies within the range 0.675 $\pm$ 0.035. On the other hand, $\chi_{\text{max}}$ remains $\gtrsim$ 0.16 for our chosen range of densities and choice of EoSs. Therefore, we choose sequences for $\chi$ values of 0.2, 0.4, 0.6, and the Keplerian sequence. It is important to note that the Keplerian sequence does not represent a fixed value of $\chi$ but rather a value which varies between 0.64 and 0.71. This slight variation in $\chi_{\text{max}}$ at saturation depends mostly on the EoSs. For instance, the highest value of 0.71 is obtained using the Nucleonic EoS DD2, whereas the inclusion of heavy baryon degrees of freedom reduces the saturation value slightly. Consideration of quark core further reduces the $\chi_{\text{max}}$ value at saturation, and the dashed line curves in fig \ref{fig2b_chi} is indicative of the fact. 

The mass-equatorial radius relations for HSs using DD2Q EoS with the chosen spin sequences, including the slowly rotating one, are shown in Fig. \ref{fig2a_M-R}. The mass-radius trend remains the same for all the spin sequences. However, for a chosen central density, the gravitating mass increases due to the support from centrifugal force due to rotation. Therefore, the spin sequences with larger $\chi$ obtain larger mass than their slowly rotating counterparts. The spin-induced deformation leads to an oblate shape of the star, resulting in radii increasing with spin. Next, we investigate the quasi-universal relations for these spin sequences. 

Our calculated results for $\overline{I}$-C-$\overline{Q}$ relations are presented in the upper panels of Figs. \ref{fig8_I-C}, \ref{fig9_Q-C}, and \ref{fig10_I-Q} for all the chosen values of $\chi$. The relations are determined for composition (a) and (b) and presented along with the computed results for composition (b), similar to the case of slowly rotating stars. Using Eq. \ref{eqn11}, we develop $\overline{I}$-C and $\overline{Q}$-C relations following the same procedure as $\chi = $ 10$^{-3}$ case. Similarly, we obtain the $\overline{I}$-$\overline{Q}$ relation using Eq. \ref{eqn10} and calculate the corresponding fractional deviations of calculated results. All the lower panels of Figs. \ref{fig8_I-C}, \ref{fig9_Q-C} \& \ref{fig10_I-Q} show relative errors of the aforementioned relations. It is worth mentioning the fact that for all the chosen spin values, the relative errors slightly increase with the spin parameter $\chi$. The parameters of $\overline{I}$-C, $\overline{Q}$-C, $\overline{I}$-$\overline{Q}$ relations are summarized in table \ref{table3}. The asymptotic standard error of the relation parameters turns out to be higher for HSs and increases further with the spin parameter value. Thus, the addition of deconfined quark degree of freedom in the core composition of rapidly rotating stars adds to the variability of the quasi-universal relations.

Figs. \ref{fig8_I-C}, \ref{fig9_Q-C} \& \ref{fig10_I-Q} (top) also show $\overline{I}$-C-$\overline{Q}$ relations for the maximally rapidly rotating sequence. Similar to other sequences with fixed $\chi$ values, the $\overline{I}$-C-$\overline{Q}$ relations are explored using Eq. \ref{eqn10} and Eq. \ref{eqn11} with both compositions (a) and (b). Relative errors between computed results and estimated universal relations obtained for composition (b) are presented in the lower panels of the figures. Though the $\chi$ value for this particular sequence varies from one EoS to another, the fractional deviations for the $\overline{I}$-C, $\overline{Q}$-C and $\overline{I}$-$\overline{Q}$ relations remain comparable to those for other spin sequences. The fitting parameters for this sequence are summarized in table \ref{table3} with the caption `Kep'. The values of $\chi_{\text{red}}^2$ are also given in table \ref{table3} for all the sequences. For the Keplerian sequence, we also observe a slight variation in the parameters of universality relations like other spin sequences. Thus, the universal relations between NS parameters are extendable to the extent of maximally rotating sequences with arbitrary general core composition.

In the existing literature, a few relations for rapidly rotating stars are investigated. For instance, ref \cite{Riahi_2019_yNGWj} obtains eos-insensitive relations between rotating frequency, mass and radius of keplerian configuration and radius of static configuration modelling maximally rapidly rotating stars NSs with 12 different hadron EoSs. Ref \cite{Khosravi_Largani_2022_cvIJr} obtains $\overline{I}$-C relation for keplerian sequences of rotating zero temperature and finite isentropic stars. Their EoS model considers hadron matter with few of the exotic degrees of freedom taken into account and quark matter treated within the description of MIT Bag \cite{Fischer_2011_DbzLV}, vector-interaction-enhanced bag model \cite{Kl_hn_2015_lo3ys}, String-flip models \cite{Bastian_2018_K1F9o} and NJL Lagrangian based model \cite{Klevansky_1992_2oEgq, Beni__2015_PdKX3, Baym_2018_fVTXR}.

Authors in ref \cite{Breu_2016_WqIRD} model NSs with 28 different EoSs, mostly considering nuclear degrees of freedom, to obtain the $\overline{I}$-C relation. The result of this study shows a relative error of $\sim$ 20\% for a rapidly rotating sequence of $\chi =$ 0.6. Our obtained relations with core composition variability further add to the literature. In our calculation, we find the relations using the method of ref \cite{Yagi_2013_cPzvP}. The relative errors from our fit formulas increase with $\chi$, and we obtain the relative difference for $\overline{I}$-C relation to be $\lesssim$ 10\% even for $\chi =$ 0.6.

$\overline{I}$-$\overline{Q}$ relation is obtained for rapidly rotating NSs implementing a fit function of the form $\overline{I} = \overline{I}\left( \overline{Q}, \chi \right)$ using hadron EoSs in refs \cite{Pappas_2014_TIiDy, Chakrabarti_2014_kBsdN} with relative errors of $\sim$ 1\% for arbitrary rotation. A similar fitting method is used within a machine-learning technique to obtain EoS-insensitive relations between several integral quantities of NSs and HSs in ref \cite{Papigkiotis_2023_a8kst}. Thus $\overline{I} = \overline{I}\left( \overline{Q}, \chi \right)$ relation is obtained within $\sim$ 2\% deviation for an ensemble of EoSs composed of nucleons, nucleons with hyperonic degrees of freedom, and nucleons with deconfined quark degrees of freedom. In the present work, we obtain $\overline{I} -\overline{Q}$ relation for compact stars with arbitrary core composition and along fixed values of the spin parameter. We observe that the deviation from fit functions remains within $\sim$ 3-4\% for the mentioned relation.

\section{Summary and Conclusion}
\label{conclusion}
The study of the EoS-insensitive relations finds particular importance in the literature \cite{Rezzolla_2018_Gl7Zu, Kumar_2019_PdDCq}. For instance, figs. \ref{fig8_I-C}, \ref{fig9_Q-C}, and \ref{fig10_I-Q} provide a basis for determining the radius of a pulsar when the mass and either $\overline{I}$ or $\overline{Q}$ are accurately measured. Mass measurements in neutron star binaries enable the derivation of pulsar radii based on the definitions of $\overline{I}$ and $\overline{Q}$. However, the degeneracy between $\overline{I}$ and $\overline{Q}$ is significant for pulsars with low spin or high compactness, posing a considerable challenge in independently measuring pulsar radii using either moment of inertia or quadrupole moment. In this work, we address the universal relations for NSs, encompassing a comprehensive range of exotic degrees of freedom in nuclear matter at extreme densities. For slowly rotating NSs, we extend the C-$\overline{\Lambda}$ relation of refs \cite{Maselli_2013_KEJBr, Yagi_2017_uWwLw, Raduta_2020_YKkHx, Yeung_2021_dc0JE, Burikham_2022_Kq9sd, Carlson_2023_0TLyw, Li_2023_wnUwG} with a very general composition of the core. I-Love-Q relations for slowly rotating NSs are studied in the literature, including both nucleonic and heavy baryon degrees of freedom \cite{Baub_ck_2013_gJPHq, Urbanec_2013_wNmiD, Yeung_2021_dc0JE}. We generalize the existing relations by adding the quark degrees of freedom to complement the findings of refs \cite{Suleiman_2024_7cKmd, Legred_2024_WRfSP}, which use a large agnostic set of EoSs to mimic the hybrid EoSs. For rapidly rotating NSs, we also extend the existing $\overline{I}$-C relation of refs \cite{Breu_2016_WqIRD, Khosravi_Largani_2022_cvIJr, Li_2023_wnUwG} and $\overline{I}$-$\overline{Q}$ relation of refs \cite{Pappas_2014_TIiDy, Chakrabarti_2014_kBsdN, Papigkiotis_2023_a8kst} to allow almost all the exotic degrees of freedom inside the core. 

We include nucleon and heavy baryon degrees of freedom in our chosen set of EoSs. The EoSs are constructed by calibrating nucleonic sector couplings through nuclear phenomenology and heavy baryon couplings in the scalar-meson sector via symmetric nuclear matter potential fittings. These EoS models align with constraints from nuclear physics experiments and observations of NSs, particularly their radii and masses derived from NICER observations. Then, we consider the possibility of deconfined and interacting quark matter inside the core of the NS. Owing to Lattice QCD studies of the thermodynamics of hot quark-gluon plasma demonstrating the importance of considering the interactions, the description of interacting quark matter at zero temperature and sufficiently high densities is constructed within the framework of pQCD. The quark EoS and the hadronic EoSs are smoothly joined using the Maxwell construction of phase transition. The hybrid EoSs are able to produce stable and connected branches of HSs.

We investigate the measurable global properties of NSs, including mass, radius, tidal deformability, moment of inertia, and quadrupole moment for isolated stars. We summarize the effect of incorporating quark core along with different exotic phases of hadronic matter on C-$\overline{\Lambda}$ and $\overline{I}$-$\overline{\Lambda}$-$\overline{Q}$ relations for slowly rotating stars in section \ref{RD_slowNS}. We also compare our results with existing relations for slowly rotating NSs. Then, we utilize the almost EoS-independent nature of C-$\overline{\Lambda}$ relation to obtain $\overline{I}$-C-$\overline{Q}$ quasi-universal relations for rapidly rotating NSs with general composition of core. Relations for rapidly rotating NSs are presented and discussed in comparison to already existing literature. Our results show that the parameters of fit formulae have higher values of uncertainty, in general, while including deconfined quark degrees of freedom. Thus, the variation in core composition affects the accuracy of the obtained relations. The accuracy of the fit further degrades with the spin parameter.

Our derived relations represent updated versions of those already existing in the literature, tailored to EoS collections with a specific emphasis on exotic degrees of freedom. These obtained relations facilitate EoS-insensitive estimations of measurable quantities of NSs. Future astronomical observations, coupled with these relations, hold great promise for enhancing and broadening our understanding of compressed matter at astrophysical sites.

\twocolumngrid
\bibliography{pbib}
\end{document}